\documentclass[aps,twocolumn,prx,floatfix,showpacs,lengthcheck,citeautoscript,superscriptaddress,longbibliography]{revtex4-1}

\usepackage[none]{hyphenat}
\usepackage{amssymb} 
\usepackage{amsmath}
\usepackage{graphicx}
\usepackage{bm}
\usepackage{times}
\usepackage{color,soul}
\usepackage[dvipsnames,svgnames,table]{xcolor}
\usepackage{hyperref}
\usepackage{enumitem}
\usepackage[english]{babel}
\usepackage[caption=false]{subfig}
\usepackage{braket}
\usepackage{scalerel}

\usepackage{mathrsfs}
\usepackage{MnSymbol}
\usepackage{lipsum}

\hypersetup{
pdfstartview={FitH},% fits the width of the page to the window
colorlinks=true,    % false: boxed links; true: colored links
linkcolor=NavyBlue, % color of internal links
citecolor=Blue,   % color of links to bibliography
filecolor=NavyBlue, % color of file links
urlcolor=NavyBlue   % color of external links
}

\begin{document}

\title{Dynamical phases transitions in periodically driven Bardeen-Cooper-Schrieffer systems
}

\author{H. P. {Ojeda Collado}}
\email{hector.pablo.ojedacollado@roma1.infn.it}
\affiliation{ISC-CNR and Department of Physics, Sapienza University of Rome, Piazzale Aldo Moro 2, I-00185, Rome, Italy}
\author{Gonzalo Usaj}
\affiliation{Centro At{\'{o}}mico Bariloche and Instituto Balseiro,
Comisi\'on Nacional de Energ\'{\i}a At\'omica (CNEA)--Universidad Nacional de Cuyo (UNCUYO), 8400 Bariloche, Argentina}
\affiliation{Instituto de Nanociencia y Nanotecnolog\'{i}a (INN), Consejo Nacional de Investigaciones Cient\'{\i}ficas y T\'ecnicas (CONICET)--CNEA, 8400 Bariloche, Argentina}
\author{C. A. Balseiro}
%\email[Corresponding author: ]{balseiro@cab.cnea.gov.ar}
\affiliation{Centro At{\'{o}}mico Bariloche and Instituto Balseiro,
Comisi\'on Nacional de Energ\'{\i}a At\'omica (CNEA)--Universidad Nacional de Cuyo (UNCUYO), 8400 Bariloche, Argentina}
\affiliation{Instituto de Nanociencia y Nanotecnolog\'{i}a (INN), Consejo Nacional de Investigaciones Cient\'{\i}ficas y T\'ecnicas (CONICET)--CNEA, 8400 Bariloche, Argentina}
\author{Dami{\'{a}}n H. Zanette}
\affiliation{Centro At{\'{o}}mico Bariloche and Instituto Balseiro,
Comisi\'on Nacional de Energ\'{\i}a At\'omica (CNEA)--Universidad Nacional de Cuyo (UNCUYO), 8400 Bariloche, Argentina}
\affiliation{Consejo Nacional de Investigaciones Cient\'{\i}ficas y T\'ecnicas (CONICET), Argentina}
\author{Jos\'{e} Lorenzana}
\email{jose.lorenzana@cnr.it}
\affiliation{ISC-CNR and Department of Physics, Sapienza University of Rome, Piazzale Aldo Moro 2, I-00185, Rome, Italy}

\begin{abstract}
  We present a systematic study of the dynamical phase diagram of a periodically driven BCS system as a function of drive strength and frequency. 
  Three different driving mechanism are considered and compared:  oscillating density of states,  oscillating pairing interaction and oscillating external paring field. We identify the locus in parameter space of
parametric resonances and four dynamical phases: Rabi-Higgs,  gapless, synchronized Higgs  and time-crystal.  We demonstrate that the main features of the phase diagram are quite robust to different driving protocols and discuss the order of the transitions. By mapping the BCS problem to a collection of nonlinear and interacting classical oscillators, we shed light on the origin of time-crsytalline phases and parametric resonances appearing for subgap excitations. 
%Taking advantage of the first order phase transition, we propose a driving protocol giving rise to hysteretic behaviour
\end{abstract}

\date{\today}
\maketitle

%%%%%%%%%%%%%%%%%%%%%%%%%%%%%%%%%%%%%%%
\section{Introduction}
%%%%%%%%%%%%%%%%%%%%%%%%%%%%%%%%%%%%%%%
%The French mathematician Floquet was the first to obtain rigorous results\cite~{floquet1883} on dynamical systems subject to a periodic drive.
%The manipulation of many-body systems by periodic drives is termed ``Floquet engineering''.

The manipulation of many-body systems by periodic drives, usually referred to as 
``Floquet engineering'', has become a powerful tool to control  properties of materials~\cite{Oka2019,Bukov2015,Claeys2019,Meinert2016}.
Floquet engineering has benefit from the tremendous advances in laser technologies and the advent of highly controllable systems~\cite{Eisert2015} like ultracold atomic gases in optical traps~\cite{Behrle2018,Chin2010,Eckardt2017} or   cavities~\cite{Muniz2020,Lewis-Swan2021,Norcia2018}, ion chains~\cite{Zhang2017c} and nuclear spins.
Some experimental demonstrations include, quantum control of magnetism~\cite{Gorg2018}, topology~\cite{mclver2019}, electron-phonon interactions~\cite{Borroni2017}
and broken symmetry phases such as superconductivity~\cite{Fausti2011,Mitrano2016}.

Superconductors are also one of the most popular platforms for quantum technologies. Quantum devices require the manipulation of the out-of-equilibrium system  for as long as possible without loosing coherence because of coupling to the environment. Thus, in the last years a large effort has been done to improve the materials and devices to increase the energy relaxation time and the coherent dynamics~\cite{Mannila2022,Catelani2022,Saira2012}.

Superconducting and ultracold atomic superfluid condensates present a
unique opportunity to study many-body Floquet effects~\cite{OjedaCollado2018,Claassen2019,OjedaCollado2020,Homann2020,Yang2020a,Yang2020,OjedaCollado2021,Pena2022,Pena2022a}. Because of a gap in the excitation spectrum, these systems tend to have long relaxation times. In addition, in weak-coupling, the dynamics can be described by a mean-field like Hamiltonian with effective all-to-all interactions. Both effects contribute to provide a large window of time where energy relaxation processes are suppressed and the out-of-equilibrium dynamics can be studied. 
Furthermore, all-to-all interacting systems have aroused interest in the context of mean-field time-crystals~\cite{Natsheh2021,Natsheh2021b,Else2020,Lazarides2020,OjedaCollado2021}.

Notwithstanding all this growing interest, Floquet engineering in superconducting or superfluid condensates  has not been addressed until recently~\cite{Sentef2017,OjedaCollado2018,Kennes2019,OjedaCollado2019,OjedaCollado2020,Homann2020,DelaTorre2021,OjedaCollado2021,Buzzi2020,Puviani2021,Lyu2022}. For periodically driven BCS systems, Rabi-Higgs oscillations~\cite{OjedaCollado2018}, parametric resonances and Floquet time-crystal phases~\cite{Homann2020,OjedaCollado2021} have been demonstrated by considering a periodic time-dependent pairing interaction $\lambda (t)$.

Despite this progress, several questions remain open. Different dynamical phases have been identified~\cite{OjedaCollado2018,OjedaCollado2019,OjedaCollado2020,Homann2020} and a partial dynamical phase diagram has been presented for the driven BCS system in Ref.~\cite{OjedaCollado2021}. On the other hand,  the order of the transition has not been discussed.
Also, so far studies have concentrated on a driving mechanism in which the interaction parameter $\lambda$ is time-dependent ($\lambda$-driving). However, 
it is also possible to envisage that the  density-of-states (DOS) could be time dependent (DOS-driving). 

%other mechanisms are possible as density-of-states driving (DOS-driving) and
%driving by an external pairing field. 

Here we present a systematic study of the dynamical phase diagram of a driven BCS system including driving frequencies such that $\hbar \omega$ lies below and above the gap and a large range of drive amplitudes and both $\lambda-$ and DOS-driving mechanisms. The DOS-driving protocol is relevant for ultracold atoms setups as well as in condensed-matter systems where the electrons can couple to an electromagnetic field in the THz regime.
% Pairing field driving can be relevant in multiband systems (where one band can drive a second band).
In addition, a systematic comparison of driving mechanisms allow to separate universal features from mechanism-dependent details.

We show that the phase diagram is, in general, surprisingly rich with at least four dynamical phases (Rabi-Higgs, gapless, synchronized-Higgs and time-crystal)
ubiquitously appearing for both driving protocols. Dynamical phase transitions (DPTs) are analyzed in detail and we demonstrate the existence of first and second-order like phase transitions. We analyze the parametric resonances discovered before~\cite{OjedaCollado2021} and discuss their origin in the context of the mapping to a classical dynamical system. In order to clarify the essential ingredients leading to parametric resonances, we compare the phase diagram for $\lambda-$ and DOS-driving with the one corresponding to an external pairing field (third driving mechanism). Also, to highlight the relevance of the many-body interactions in the emergence of parametric resonances, we compare these phases diagrams to the one obtained in the case that the self-consistency of the BCS order parameter is neglected. 

The paper is organized as follows: Section~\ref{sec:peri-driv-bcs}  introduces the model and the methods used. Sec.~\ref{sec:phase-diagrams} presents the dynamical phase diagrams. Section~\ref{sec:dyna} discusses the dynamics in each phase. Section~\ref{sec:dynam-phase-trans} analyzes the order of the transitions. In Sec.~\ref{sec:mapclas} we present the mapping to a classical system of non-linear oscillators. Finally, in Sec.~\ref{sec:conclusions} we present our conclusions.

% After the demonstration of Floquet time-crystal phases emerging inside the parametric resonances a still open question is what is the origin of such behavior. Here we shed light on the essential ingredients to obtain parametric resonances in periodically driven BCS systems by mapping the problem to a collection of nonlinear classical oscillators.

%%%%%%%%%%%%%%%%%%%%%%%%%%%%%%%%%%%%%%%%%%%%%%%%%%%%%%%%%%%%%
\section{Periodically driven BCS model}\label{sec:peri-driv-bcs}
%%%%%%%%%%%%%%%%%%%%%%%%%%%%%%%%%%%%%%%%%%%%%%%%%%%%%%%%%%%%%%%
\subsection{The pseudospin model}

We consider the following time-dependent BCS Hamiltonian written in terms of  
Anderson pseudospins~\cite{Anderson1958},
%%%%%%%%%%%%
\begin{equation}\label{eq:ham}
\hat{H}_{\mathrm{BCS}}=-2\sum_{\bm{k}} \xi_{\bm{k}}(t)\hat{S}_{\bm{k}}^{z}-\lambda(t)\sum_{\bm{k},\bm{k}'}\hat{S}_{\bm{k}}^{+}\hat{S}_{\bm{k}'}^{-}\,.
\end{equation} 
%%%%%%%%%%%%
Here, $\xi_{\bm{k}}=\varepsilon_{\bm{k}}-\mu$ measures the energy of the fermions ($\varepsilon_{\bm{k}}$) from the Fermi level $\mu$ and $\lambda$ is the pairing interaction. Either 
$\xi_{\bm{k}}(t)$ or $\lambda(t)$ is taken as time-dependent. In the first case, for a uniform rescaling of the fermionic band, we can consider the
DOS itself $\nu$ to be time-dependent (DOS-driving) while the second case defines $\lambda$-driving.  More details of the protocols will be given in the next subsection.

The $\frac{1}{2}$-pseudospin operators are given in terms of fermionic operators as, 
%%%%%%%%%%%%%%%%%%%
\begin{eqnarray}
\nonumber
\hat{S}_{\bm{k}}^{x}&=&\frac{1}{2}\left(\hat{c}_{\bm{k}\uparrow}^{\dagger}\hat{c}_{-\bm{k}\downarrow}^{\dagger}+\hat{c}_{-\bm{k}\downarrow}^{}\hat{c}_{\bm{k}\uparrow}^{}\right)\,, \\ \label{eq:pesudospins}
\hat{S}_{\bm{k}}^{y}&=&\frac{1}{2i}\left(\hat{c}_{\bm{k}\uparrow}^{\dagger}\hat{c}_{-\bm{k}\downarrow}^{\dagger}-\hat{c}_{-\bm{k}\downarrow}^{}\hat{c}_{\bm{k}\uparrow}^{}\right)\,, \\
\hat{S}_{\bm{k}}^{z}&=&\frac{1}{2}\left(1-\hat{c}_{\bm{k}\uparrow}^{\dagger}\hat{c}_{\bm{k}\uparrow}^{}-\hat{c}_{-\bm{k}\downarrow}^{\dagger}\hat{c}_{-\bm{k}\downarrow}^{}\right)\nonumber
\,,
\end{eqnarray}
%%%%%%%%%%%%%%%%%
and $\hat{c}^\dagger_{\bm{ k}\sigma}$ ($\hat{c}_{\bm{k}\sigma}$) is the usual creation (annihilation) operator for fermions with momentum $\bm{k}$ and spin $\sigma$. The operator $\hat{S}_{\bm{k}}^{\pm}\equiv \hat{S}_{\bm{k}}^{x}\pm i\hat{S}_{\bm{k}}^{y}$ creates or annihilates a Cooper pair $(\bm{k},-\bm{k})$.

Due to the all-to-all interaction, assumed in the second term of Eq.~(\ref{eq:ham}), one can use a time-dependent mean-field treatment ~\cite{Barankov2004,Yuzbashyan2005,Yuzbashyan2005a,Barankov2006,Barankov2006a,Yuzbashyan2006,Yuzbashyan2006a,Chou2017,Hannibal2015a,Yuzbashyan2015,Scaramazza2019,Seibold2021,Collado2022} which yields 
the {\em exact} dynamics in the thermodynamic limit. 
The BCS mean-field Hamiltonian can be written as,
%%%%%%%%%%%%%%%%%
\begin{equation}\label{eq:hamMF}
\hat{H}_{\mathrm{MF}}=-\sum_{\bm{k}}\hat{\bm{S}}_{\bm{k}}\cdot\bm{b}_{\bm{k}}.
\end{equation}
%%%%%%%%%%%%%%%%%
where, $\bm{b}_{\bm{k}}\left(t\right)=(2\Delta\left(t\right),0,2\xi_{\bm{k}})$ is the mean-field acting on the  the $\frac{1}{2}$-pseudospin operator $\hat{\bm{S}}_{\bm{k}}=(\hat{S}_{\bm{k}}^{x},\hat{S}_{\bm{k}}^{y},\hat{S}_{\bm{k}}^{z})$. The pseudomagnetic field $\bm{b}_{\bm{k}}$ has to be obtained in a self-consistent manner during the dynamics.

Without loss of generality, we consider that the equilibrium superconducting order parameter $\Delta_0$ is real.
We will assume this 
remains valid over time and show below that this is indeed the case 
because of the electron-hole symmetry. 

The real part of the instantaneous BCS order parameter is given by 
%%%%%%%%%%%%%%%%%%%%%%%
\begin{equation}
\label{eq:deltat}
\Delta(t)=\lambda(t)\sum_{\bm{k}} S_{\bm{k}}^{x}\,,
\end{equation}
%%%%%%%%%%%%%%%%%%%%%%%%%
where $S_{\bm{k}}^{x}$, without hat, denotes the expectation value of the operator $\hat{S}_{\bm{k}}^{x}$ in the time-dependent BCS state. Hereon, we will use this notation for all pseudospins components.

In practice, 
since the pseudomagnetic field depends on $\bm{k}$ only through $\xi_{\bm{k}}$,
rather than solving the equations for each  $\bm{k}$ we  solved the equations for a generic 
DOS converting the sums into integrals over the fermionic energy $\xi$,
%%%%%%%%%%%%%%%%%%%%%%%
\begin{equation}
\label{eq:deltatdos}
\Delta(t)=\lambda(t)\int d\xi \nu(\xi)S^{x}(\xi)\,,
\end{equation}
%%%%%%%%%%%%%%%%%%%%%%%%%
with $S^{x}(\xi_{\bm{k}})\equiv S^{x}_{\bm{k}}$ and similar for the other components---we shall use  $S^{x}_{\bm{k}}$ and
$S^{x}(\xi_{\bm{k}})$ interchangeably, keeping in mind that in actual computations the $\xi$-dependent form was used.  

At equilibrium, in the absence of periodic perturbations, the $\frac{1}{2}$-pseudospins align in the direction of their local fields $\bm{b}_{\bm{k}}^0=(2\Delta_{0},0,2\xi_{\bm{k}})$ in order to minimize the system's energy [described by  Eq.~(\ref{eq:hamMF})]. This corresponds to the zero-temperature paired ground state in which the pseudospin texture (the expectation value of pseudospin operators as a function of momentum $\bm{k}$) is given by
%%%%%%%%%%%%%%%%%%
\begin{equation}
\label{eq:sequilib} 
S_{\bm{k}}^{x,0} =\frac{\Delta_{0}}{2\sqrt{\xi_{\bm{k}}^{2}+\Delta_{0}^{2}}},\: S_{\bm{k}}^{y,0} =0, \: S_{\bm{k}}^{z,0} =\frac{\xi_{\bm{k}}}{2\sqrt{\xi_{\bm{k}}^{2}+\Delta_{0}^{2}}}.
\end{equation}
%%%%%%%%%%%%%%%%%
Such pseudospin texture is used as initial condition and once the pairing interaction or the DOS is modulated in time, the expectation values of the pseudospins evolve obeying a Bloch-like equation of motion
%%%%%%%%%%%%%%%%%%%%
\begin{equation}
\label{eq:eom}
\frac{d\bm{S}_{\bm{k}}}{dt}=-\bm{b}_{\bm{k}}\left(t\right)\times\bm{S}_{\bm{k}}\,,
\end{equation}
%%%%%%%%%%%%%%%%
where we set $\hbar\equiv1$.

We assume that the time dependent solutions do not spontaneously break
particle-hole symmetry.
From the equations of motion one can check that
if  $b^y(\xi)=\Delta''=0$ then since $b^x(\xi)=b^x(-\xi)=2\Delta$ and $b^z(\xi)=b^z(-\xi)$ 
the self-consistent solution preserves the following symmetries, 
\begin{eqnarray}
  \label{eq:ph}
  S^{x}(\xi)&=&S^{x}(-\xi),\nonumber\\
 S^{y}(\xi)&=&-S^{y}(-\xi),\\
 S^{z}(\xi)&=&-S^{z}(-\xi).   \nonumber
\end{eqnarray}
Indeed, the imaginary part of the order parameter is given by,
\begin{equation}
\label{eq:imdeltatdos}
\Delta''(t)=\lambda(t)\int d\xi \, \nu(\xi)\,S^{y}(\xi),
\end{equation}
which vanishes if Eq.~(\ref{eq:ph}) holds [and $\nu(\xi)=\nu(-\xi)$ as assumed].
Now, by considering $\Delta''$ at a time $t+dt$,
\begin{eqnarray}
  \label{eq:delta2}
     \Delta''(t+dt)-\Delta''(t)&=& dt\lambda(t)
    \nonumber\\
  &\times&\!\!\!    \int\! d\xi\, \nu(\xi) [b^x(\xi) S^z(\xi)-b^z(\xi) S^x(\xi)]\nonumber\\
    &=&  0.
\end{eqnarray}
This shows that the $\Delta''(t)=0$ is preserved at all times. Thus, our initial assumption and also Eqs.~(\ref{eq:ph}) are self-consistently satisfied at all times.   

\subsection{Numerical implementation}

In our computations, we consider typically $N=10^4$ pseudospins associated to equally spaced discrete energy states $\xi_{\bm{k}}$ within an energy range of $W=40\Delta_0$ around $\mu$ with an energy constant density of states $\nu$. The $N$ coupled differential equations arising from Eq.~(\ref{eq:eom}) are solved using a standard Runge-Kutta 4th-order method with a small enough $dt$ ensuring the convergence of dynamics. Some selected points in the phase diagram were also checked using an adaptive step-size Runge-Kutta method (Fehlberg method).

\subsection{Driving protocols}

In the following, we consider two different driving protocols. In the $\lambda$-driving case, the pairing interaction is taken periodic in time, as
\begin{equation} 
\lambda(t)=\lambda_0\,[1+\alpha\, \sin(\omega_d t)]\,,
\end{equation} 
while $\xi_{\bm{k}}$ does not depend on time. Here, $\lambda_0$ is the equilibrium coupling constant, $\alpha$ is the driving strength, and $\omega_d$ is the drive frequency.

In DOS-driving, we consider a time-periodic DOS with a time independent pairing interaction $\lambda_{0}$. This can be achieved with a periodic modulation of the Fermi velocity which corresponds to a change in the band structure as
\begin{equation} 
\xi_{\bm k}\left(t\right)=\xi_{{\bm k}}^{0}\left[1+\beta\sin\left(\omega_{d}t\right)\right]\,,
\end{equation}
yielding a time-dependent DOS given by
\begin{equation}
\nu\left(t\right)=\frac{\nu_{0}}{1+\beta\sin\left(\omega_{d}t\right)}\label{eq:dost}\,,
\end{equation} 
where $\nu_0$ is the constant DOS at equilibrium.
The equilibrium $T_c$ and order parameter depend on the product
$\lambda\nu$ so an adiabatic change in either parameter is equivalent. In contrast, the two protocols are rather different when the system is out of equilibrium and  produce different dynamics. DOS-driving implies that there is a momentum and time dependent pseudomagnetic field along $z$ [through $\xi_{\bm k}(t)$] which acquires $x$-components once $\Delta$ becomes time dependent. On the other hand, $\lambda-$ drive means a time-dependent pseudomagnetic field only along the $x-$direction. Possible experimental implementations of both protocols in ultracold atoms and condensed matter systems have been discussed in detail in Ref.~\cite{OjedaCollado2018}.

%In contrast with previous studies~\cite{OjedaCollado2018,OjedaCollado2020}, we will focus not only in the dynamical phases but also in the DPTs by considering a large range of amplitudes and drive frequencies. 

%%%%%%%%%%%%%%%%%%%%%%%%%%%%%%%%%%%%%%%%%%%%%%%%%%%%%
\begin{figure}[tb]
\includegraphics[width=0.5\textwidth]{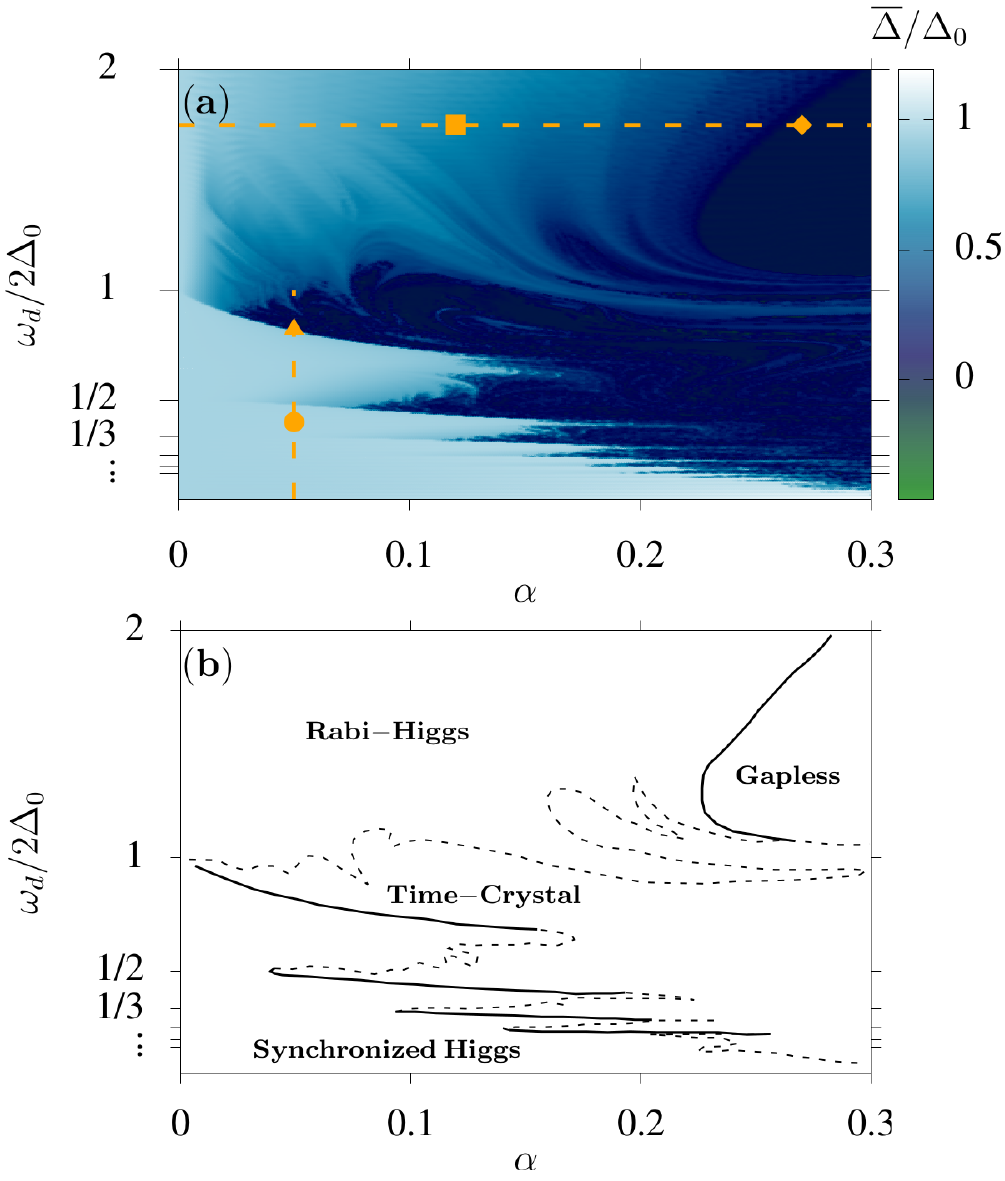}
\caption{(Color online) (a) Temporal average of superconducting order parameter $\bar{\Delta}$ as a function of amplitude and frequency of the drive, considering a $\lambda-$driving protocol. We have computed $\bar{\Delta}$ using the time window $t\Delta_0\in [0,200].$  Dashed orange lines indicate cuts to be show later (Figs.~\ref{fig:verticalcut},\ref{fig:horizontalcut}).  The orange dots indicate the parameters where the detailed dynamics 
  is displayed (Fig.~\ref{fig:lambdadyn}). They correspond to the different  dynamical phases found: Synchronized Higgs  (circle), time-crystal  (triangle), Rabi-Higgs  (square) and gapless (rhombus). (b) Schematic representation of the phase diagram showing the dominant phases in each region.  Dashed curves represent continuous  phase transitions with fractal-like boundaries  while solid lines represent first-order DPTs as shown below.} 
\label{fig:phased}
\end{figure}
%%%%%%%%%%%%%%%%%%%%%%%%%%%%%%%%%%%%%%%%%%%%%%%%%%%%%%%%

\section{Dynamical Phase Diagrams}\label{sec:phase-diagrams}
 
In this section, we  present the dynamical phase diagram with both driving methods and in a wide range of frequency and driving strengths.
Previous studies focused on $\lambda$-driving and the subgap regime~\cite{OjedaCollado2021} or specific frequencies~\cite{OjedaCollado2018,OjedaCollado2020}.

A first screening of the phase diagram can be obtained~\cite{OjedaCollado2021} using the time-averaged superconducting order parameter $\bar{\Delta}$ as dynamical order parameter.
In Fig.~\ref{fig:phased} we show a false color map of $\bar{\Delta}$ as a function of the amplitude and frequency of the drive for the $\lambda-$driving protocol. 
At first look, there are two main regions that can be easily distinguished:  in the light blue regions the average of the superconducting order parameter is near the equilibrium value ($\bar{\Delta}\approx\Delta_0$) while regions with zero order parameter average (ZOPA) appear in dark blue.
We identify four different dynamical phases within these regions, which are schematized and labeled in panel (b).  However, these need a more refined analysis to be distinguished, as explained below.

%%%%%%%%%%%%%%%%%%%%%%%%%%%%%%%%%%%%%%%%%%%%%%%%%%%%%
\begin{figure}[tb]
\includegraphics[width=0.5\textwidth]{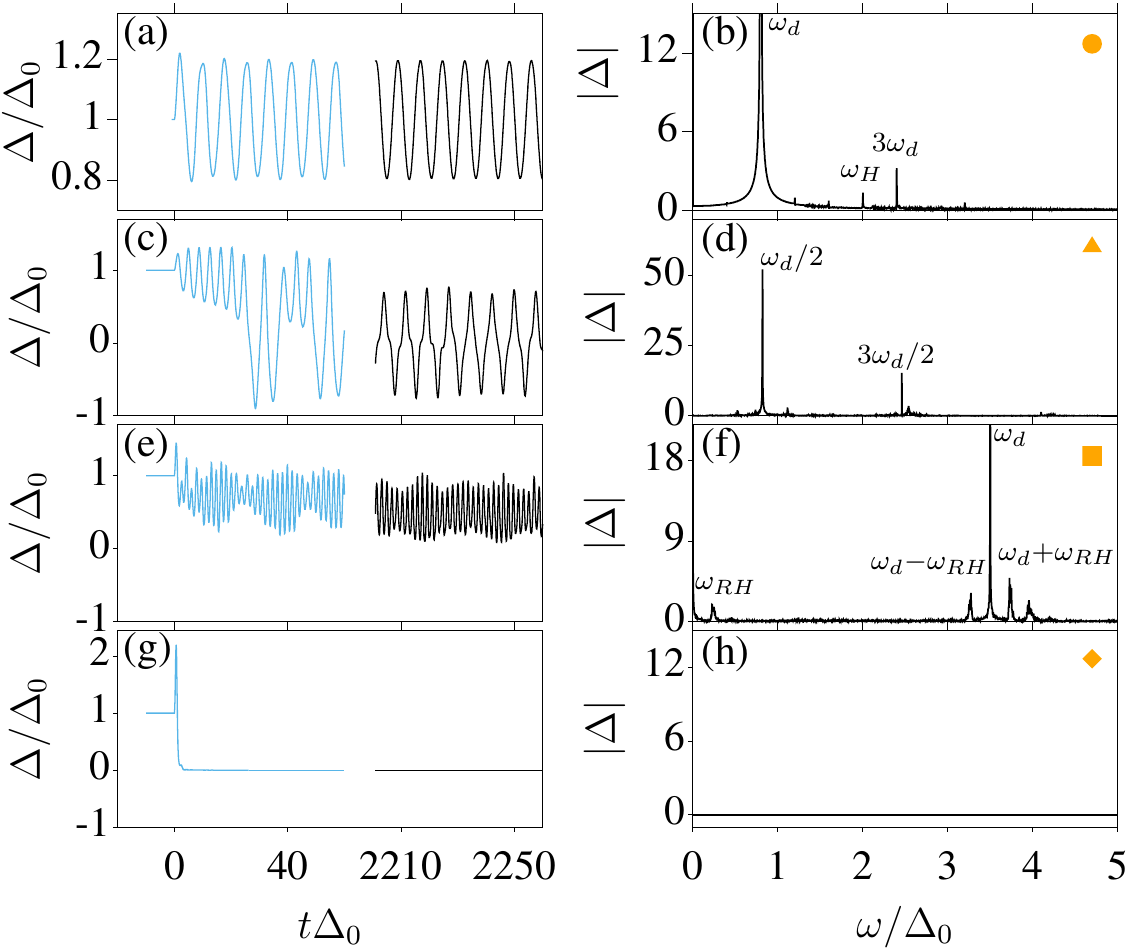}
\caption{(Color online) Representative dynamical phases for the highlighted parameters (filled dots) of Fig.~\ref{fig:phased} (a). In the left column we show the dynamics (transient in blue and steady state in black) and the FT in the corresponding right panel. For the Fourier analysis, we consider a long time window $t\Delta_0\in [200,2500]$ to obtain better resolution. (a, b) Synchronized Higgs mode phase in which the system responds with the Higgs frequency $\omega_H=2\bar{\Delta}$. (c, d) Time-crystal phase as a consequence of a time-translational symmetry breaking in which the order parameter shows period-doubling oscillations. (e, f) Rabi-Higgs mode phase, in which the superconducting order parameter oscillates with a low fundamental frequency $\omega_{RH}$. Also in this regime, the Higgs mode $\omega_H$ appears as a very low peak in the FT which is not appreciable in the present scale but in the log scale used in Fig~\ref{fig:verticalcut}. (g,h) Gapless regime where $\Delta(t)$ goes very rapidly to zero  and remains zero over time without exhibiting appreciable oscillations.} 
\label{fig:lambdadyn}
\end{figure}
%%%%%%%%%%%%%%%%%%%%%%%%%%%%%%%%%%%%%%%%%%%%%%%%%%%%%%%%

Two dynamical phases appear for subgap excitations, and two when the system is driven above the gap $\omega_d>2\Delta_0$. This rather strong distinction could be anticipated as in one case it is not possible to directly excite quasiparticles in the system (for subgap excitations in an off-resonant regime) while for $\omega_d>2\Delta_0$ it is possible.

For $\omega_d<2\Delta_0$, dark indentations or  ``Arnold tongues'' 
appear at  $\omega_d=2\Delta_0/n$, with $n$ a natural number. These are the parametric resonances reported in Ref.~\cite{OjedaCollado2021}. The phase outside the Arnold tongues, labeled ``synchronized Higgs'', is characterized by an order parameter quite close to equilibrium (light blue regions). In contrast, 
for $\omega_d>2\Delta_0$ the non-ZOPA phases are characterized by a smaller average order parameter.  Indeed,  the light blue regions are darker when $\omega_d>2\Delta_0$ than in the opposite case,  indicating  more quasiparticles excitations in the steady state.

In some regions the phase diagram has a marbled aspect indicating that, in general, different dynamical phases intermix.
However, regions with predominance of a given phase can be identified as schematized in the lower panel.

% To illustrate the dynamical phases, we choose some parameters (orange dots in Fig. 1 and red dots in Fig. 2) and show the dynamics and Fourier Transforms (FT) in Fig. 3 and 4.

An example of the time evolution and its Fourier transform (FT) for each one of the dynamical phases is shown in Fig.~\ref{fig:lambdadyn}. The orange dots allow to associate the parameters of each row with their location in the phase diagram.

%%%%%%%%%%%%%%%%%%%%%%%%%%%%%%%%%%%%%%%%%%%%%%%%%%%%%
\begin{figure}[tb]
\includegraphics[width=0.5\textwidth]{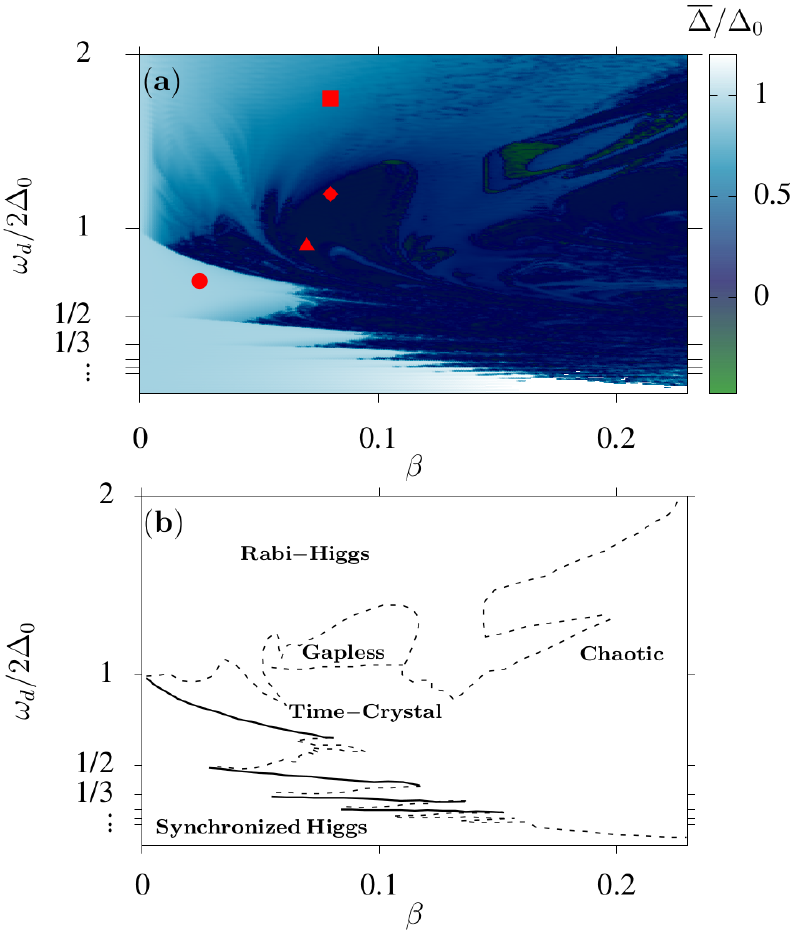}
\caption{(Color online) (a) Temporal average of superconducting order parameter $\bar{\Delta}$ as a function of amplitude and frequency of the drive considering a DOS-driving protocol. We have computed $\bar{\Delta}$ using the time window $t\Delta_0\in [0,200].$  The red dots indicate the parameters used to illustrate the more prominent dynamical phases appearing in the phase diagram: Synchronized Higgs phase (circle), time-crystal phase (triangle), Rabi-Higgs phase (square) and gapless phase (rhombus) (see App.~\ref{app:example-dynamics-dos}). (b) Schematic representation of the phase
diagram showing the dominant phases in each region. Dashed curves
represent continuous phase transitions with fractal-like boundaries
while solid lines represent first-order DPTs. In the rightmost part of the phase diagram, dynamical phases intermix in very small regions, forming a chaotic structure.} 
\label{fig:phasedDOS}
\end{figure}
%%%%%%%%%%%%%%%%%%%%%%%%%%%%%%%%%%%%%%%%%%%%%%%%%%%%%%%%

In the presence of a bath~\cite{OjedaCollado2019,OjedaCollado2020} the system
can reach thermodynamic equilibrium and linear-response theory can be applied. 
In the linear regime~\cite{Volkov1974,Cea2014,OjedaCollado2018}  the time-dependent gap parameter responds with the same frequency of the drive, so the time-translation symmetry properties of the drive are preserved.
Here, without a bath, for all the four dynamical phases, this time-translation symmetry preservation does not hold, so these are exquisitely non-linear effects which can occur as prethermal phenomena~\cite{OjedaCollado2021}.

%Except for the gapless phase, all dynamical phases show some kind of time-translation symmetry breaking. 
In the synchronized Higgs phase, illustrated in panels (a) and (b) of Fig.~\ref{fig:lambdadyn}, 
the superconducting order parameter oscillates not only with the drive frequency $\omega_d$ (and high harmonics) but also with a different and incommensurate fundamental frequency given by $\omega_H=2\bar{\Delta}$ (fundamental Higgs frequency). 
%Satellites at $\omega_H\pm\omega_d$ can also be seen in the spectrum  of panel (b). 
In the time-crystal phase [panels (c) and (d)], after a transient dynamics, period-doubling oscillations are stabilized,
indicating a subharmonic response. 
This is the typical behavior of discrete time-translational symmetry breaking displayed by Floquet   time-crystals~\cite{Else2016,Yao2017,Russomanno2017,Rovny2018,Yao2020,Else2020,Giachetti2022,Munoz-Arias2022}.
Notice that the drive frequency $\omega_d$  does not appear in the FT but  
the subharmonic response remains locked at $\omega_d/2$. This behavior persists  
under changes in the drive amplitude or frequency~\cite{OjedaCollado2021} which is the hallmark of time-crystal behavior~\cite{Yao2017,Else2020}.

For $\omega_d>2\Delta_0$ and relatively small perturbation amplitudes $\alpha$, the dynamics shows a slow modulation amplitude
on top of the fast oscillations at frequency $\omega_d$
[see Fig.~\ref{fig:lambdadyn}(e)]. This low frequency mode has been denoted as $\omega_{RH}$ in the FT (f) and corresponds to the  Rabi-Higgs mode reported in Ref.~\cite{OjedaCollado2018}. In this regime, a subset of pseudospins get synchronized and perform Rabi oscillations with a frequency proportional to the amplitude of the drive. This corresponds again to a
time-translation symmetry-breaking
subharmonic response. However, the frequency of
the mode can be tuned with the external drive, which means that the response lacks ``rigidity'' and therefore does not qualify as a time-crystal phase according to the standard definitions~\cite{Yao2017,Else2020}. 
 
%The synchronized Higgs phase, illustrated in panels (a) and (b), 
%show a strong response at the frequency of the drive and a weaker response at the Higgs mode frequency $2\bar{\Delta}$. Such mode is due to the synchronization of quasiparticles and gives the name to the phase. 
%The time-crystal phase dynamics shown in panels (c) and (d) show a response at half the frequency of the drive\cite{OjedaCollado2021}. Driving above the gap one obtains the Rabi-Higgs phase [panels (e) and (f)] characterized by the appearance of the  Rabi-Higgs mode discussed in Ref.~\cite{OjedaCollado2018}. This shows up as satellites of the main response at the driving frequency and as a distinct feature at low frequency.   
%At high driving strength one finds a  gapless phase [panels (g) and (h)].

\begin{figure}[tb]
\includegraphics[width=0.5\textwidth]{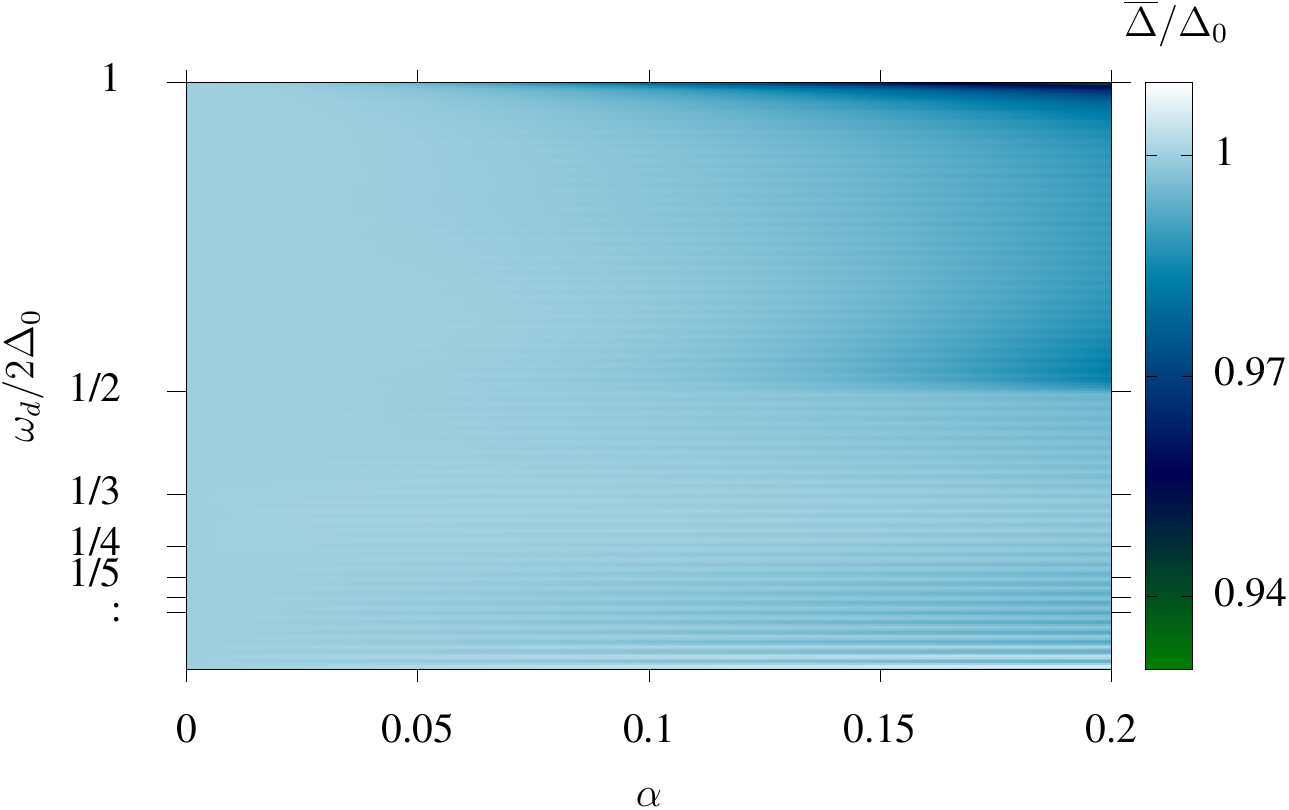}
\caption{(Color online) Dynamical phase diagram for non-interacting pseudospins.
We show the temporal average of the superconducting order parameter in the $\alpha$-$\omega_{d}$ plane for \emph{noninteracting} pseudospins subject to an oscillating (nonselfconsistent) pseudomagnetic field. In this case, parametric resonances are absent and only a weak feature  appears for subgap excitation at $\omega_d\approx\Delta_0$ and large driving amplitude. This feature becomes visible by zooming the intensity scale to a small window around $\Delta_0$ (Notice the different color bar scale respect to the other figures).  
}
\label{fig:figs2}
\end{figure}

Finally, by exciting above the gap with large drive amplitude, the system enters into a gapless regime in which the superconducting order parameter goes to zero very rapidly in time and then remains constant (g). As we shall demonstrate in the following, the fact that $\Delta(t)=0$ does not mean the absence of pairing in the system but a ``perfect" dephasing between quasiparticles. Also in this case the response does not have the same periodicity of the drive. However, instead of symmetry breaking in this case there is symmetry restoring (since the response is more symmetric in time) although in a rather trivial way.

%In the following section we discuss on the origin of this very high symmetric phase by analyzing the evolution of the pseudospin texture.
 
The real time evolution shows that for all dynamical phases there are lapses of time in which the superconducting order parameter is larger than the equilibrium value $\Delta_0$. This is more evident for the gapless regime (g) in which $\Delta (t)$ surpass $2\Delta_0$ at very short times before getting down to zero. However, this value is still below the increase we could expect by considering an adiabatic evolution~\cite{Seibold2021} where we use the instantaneous DOS or pairing interaction in the equilibrium gap equation. So this effect is rather trivial and should not be confused with dynamically induced superconductivity.

In Fig.~\ref{fig:phasedDOS} we present the same phase diagram (a) and the corresponding sketch (b) but now by considering the DOS-driving protocol. There are many common characteristics by comparing with Fig.~\ref{fig:phased}. In particular, one sees that parametric resonances for $\omega_d<2\Delta_0$ are robust features that naturally emerge independently of the protocol details.
Indeed, they appear both for periodic drive acting only along the pseudomagnetic field $x-$direction ($\lambda-$driving) or  acting along $x$ and $z-$axis at the same time (DOS-driving case). On the other hand, we show that the same dynamical phases appear with some difference in the details of  the regions of stability. In contrast to the phase diagram for the $\lambda$ driving case, here, for large perturbation amplitudes ($\beta\gtrsim 0.15$), the dynamical phase diagram becomes more chaotic where all dynamical phases practically coexist in small regions.

For completeness, we show the different dynamics at the red points in Appendix~\ref{app:example-dynamics-dos}.

\subsection{The crucial role of interactions}
\label{sec:import-inter}
The classical parametric oscillator with one degree of freedom~\cite{Landau1976}  is the simplest mechanical system to show a subharmonic response, a key ingredient of time-crystal behavior. However, time-crystals are defined also by their many-body nature. Thus, a successful way to build time-crystals is by making several parametric oscillators to interact~\cite{Heugel2019,Nicolaou2021}.

The dynamics of a single pseudospin in an external magnetic field is governed by Bloch equations, which can describe non trivial phenomena as Rabi oscillations. In  analogy with the above systems, one can wonder if the subharmonic response is already built before interactions are switch on. 
To check for this, we solve the EOM [Eq.~(\ref{eq:eom})] with a non-selfconsistent pseudomagnetic field,  
\begin{equation}
\bm{b_{\bm k}}\left(t\right)=2\left(\Delta_{0}\left[1+\alpha\sin\left(\omega_{d}t\right)\right],0,\xi_{\bm k}\right),\label{eq:magfield}
\end{equation}
starting with the equilibrium initial condition [pseudospins texture of Eq.~\eqref{eq:sequilib}].

In this case, the phase diagram becomes trivial without any visible Arnold tongues as shown in Fig.~\ref{fig:figs2} where we report a false colour plot of the temporal average of the order parameter defined here as, 
\begin{equation}
\bar{\Delta}=\lambda_0 \sum_{\bm{k}} \bar{S_{\bm{k}}^{x}}.
\end{equation}
It shows that differently from Refs.~\cite{Heugel2019,Nicolaou2021} the subharmonic response is not built in the elementary constituents but it is an emergent phenomenon which appears only after fully taking into account the quasiparticle interactions in the system.  We will come back to this problem in 
Sec.~\ref{sec:mapclas} where we will show how the system can be mapped to a collection of highly  non-linear oscillators.

\begin{figure*}[tbh]
\includegraphics[width=0.8\textwidth]{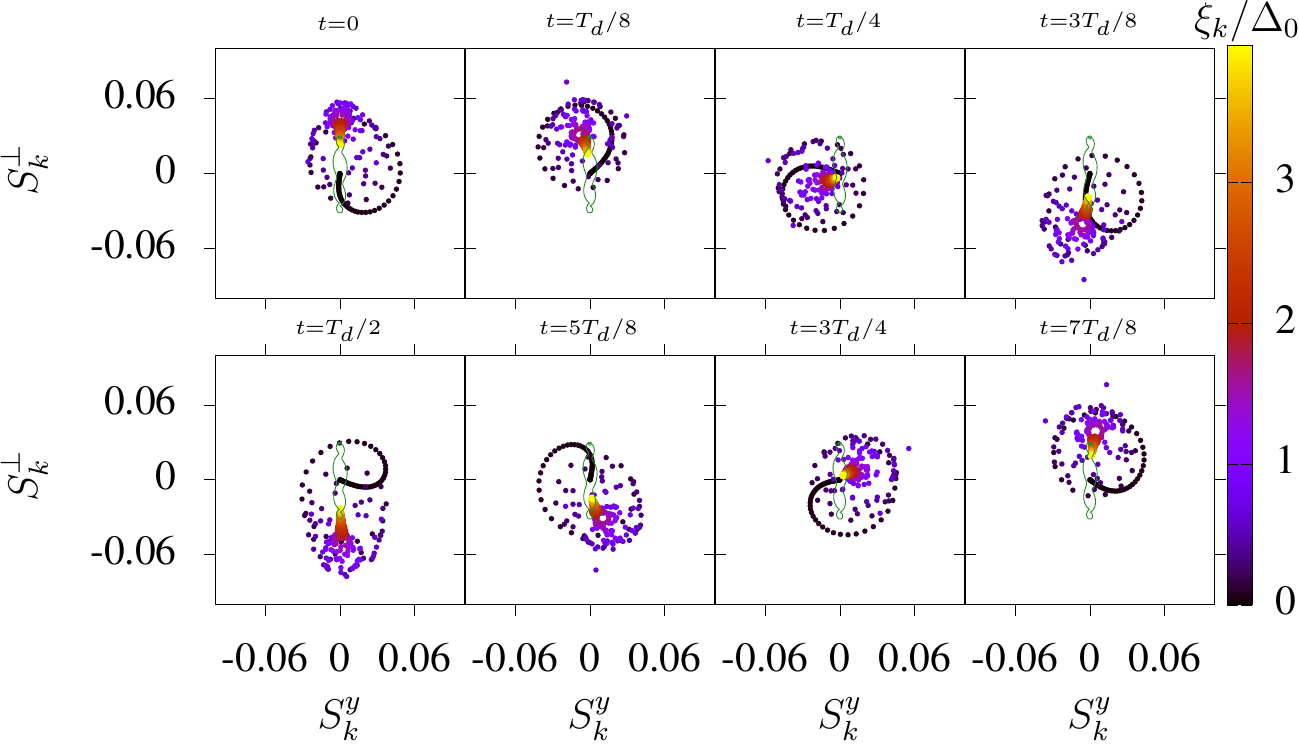}
\caption{(Color online) We show snapshots, during a time window $T_d$, of pseudospins dynamics in the synchronized Higgs mode phase ($\alpha=0.05$ and $\omega_d=0.8\Delta_0$ corresponding with the parameters used in Fig.~\ref{fig:lambdadyn} (a,b)). On the top of panels, we indicate the time. A  pseudospin trajectory for $\xi_k\simeq 3 \Delta_0$ is shown in green. See also Supplementary Video~1.}
\label{fig:snapshotHiggs}
\end{figure*}

%%%%%%%%%%%%%%%%%%%%%%%%%%%%%%%%%%%%%%%%
\section{Typical pseudospins trajectories for each dynamical phase}
\label{sec:dyna}
%%%%%%%%%%%%%%%%%%%%%%%
A more refined characterization of the dynamical phases can be obtained by studying the response resolved for each individual pseudospin. Here we show results for the $\lambda-$driving case but DOS-driving yields 
similar results.

In some cases, the dynamics is more easily analysed in terms of longitudinal and transverse components with respect to the direction of each  pseudospin
at equilibrium (i.e. without drive). 
Thus, we define a pseudospin dependent reference frame, 
hereafter the equilibrium Larmor frame (ELF), introducing the versor along the equilibrium direction, 
\begin{equation}
\hat{ \bm e}^{\parallel}_{\bm k}=2{\bm S}_{\bm k}^0=\frac2{\omega_{\bm k}}(\Delta_0,0,\xi_{\bm k}),
\end{equation}
and two transverse directions, 
  \begin{eqnarray}
    \hat{ \bm y}&=&  (0,1,0),\\
     \hat{ \bm e}^{\perp}_{\bm k}&=&  \hat{\bm y}\times \hat{ \bm e}^{\parallel}_{\bm k}=\frac2{\omega_{\bm k}}(\xi_{\bm k},0,-\Delta_0).
  \end{eqnarray}
  With these definitions the pseudospin deviations from equilibrium,  $\delta{\bm S}_{\bm{k}}\equiv{\bm S}_{\bm{k}}-{\bm S}_{\bm{k}}^0$, can be decomposed in longitudinal ($\parallel$) and transverse ($\perp$, $y$) components:  
\begin{equation}
  \label{eq:defs}
 \delta{\bm S}_{\bm{k}}=\delta S_{\bm k}^\parallel \hat{ \bm e}^{\parallel}_{\bm k} +\delta{S}_{\bm k}^\perp \hat{ \bm e}^{\perp}_{\bm k}+\delta{S}_{\bm k}^y\hat {\bm y}.
\end{equation}
Notice that since the pseudospins are normalized to length 1/2 giving two components specifies the vector up to a sign of the third component. 

%In Figs. \ref{fig:snapshotHiggs},~\ref{fig:snapshotTC},~\ref{fig:snapshotRH} and~\ref{fig:snapshotGL} we show snapshots of the pseudospins texture at different times in relevant temporal windows for the different dynamical phases using the same parametrs of Fig.~\ref{fig:lambdadyn}. 
 
%For this we defined a new ${\bm {k}}$-dependent reference frame, called  

\begin{figure*}[t]
\includegraphics[width=0.8\textwidth]{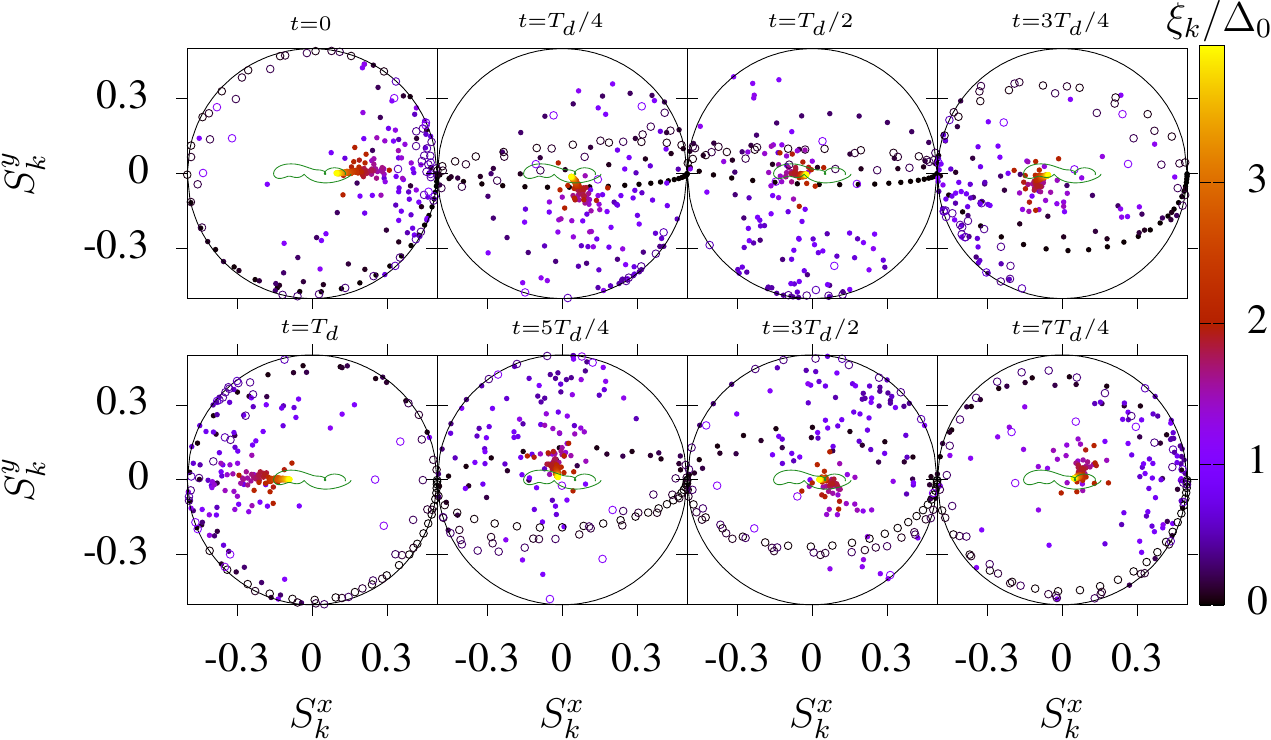}
\caption{(Color online) Dynamics in the time-crystal phase. We show snapshots, during a time window $2T_d$, of pseudospins dynamics in the time-crystal phase. Here, $\alpha=0.05$ and $\omega_d=1.64\Delta_0$ corresponding with the parameters used in Fig.~\ref{fig:lambdadyn} (c,d). On the top of panels we indicate the time. The green loop points out a typical trajectory of pseudospins (orange dots). Those pseudospins with $S^z_k>0$ ($S^z_k<0$) are indicated with filled (empty) dots. See also Supplementary Video~2.}
\label{fig:snapshotTC}
\end{figure*}

\subsection{Synchronized Higgs phase}
Figure~\ref{fig:snapshotHiggs} shows snapshots of the steady state dynamics for the synchronized Higgs phase in the ELF during a driving period, $T_d=2\pi/\omega_d$. The color  encodes the fermionic energy  $\xi_{\bm k}$. Notice that because of particle-hole symmetry [Eqs.~(\ref{eq:ph})] it is enough to show the pseudospins for $\xi_{\bm k}>0$ to specify the full texture. In this dynamical phase, the pseudospins precess very close to its equilibrium position represented by the origin. Pseudospins with quasiparticle energy $\xi_{\bm k}\simeq\Delta_0$ (purple dots),
oscillate with the drive frequency in such a way that they perform a full anticlockwise turn in a drive period $T_d$. In contrast, the low-energy pseudospins (black dots) precess more rapidly, so that by $t=3T_d/8$ they have performed more than a full turn. Their frequency $2\bar{\Delta}$ corresponds to the Higgs mode. Notice that the loop shape formed by the low-energy pseudospins (in black) preserves its form during the evolution, indicating that there is no significant dephasing. Indeed, synchronization of these pseudospins yields the main contribution to the  Higgs mode. Because of particle-hole symmetry,
the pseudospin at the Fermi level can not be excited by the drive. In general one can show that the deviation from the equilibrium position should decrease for low-energy pseudospins (black dots) as indeed observed.

\begin{figure*}[t]
\includegraphics[width=0.8\textwidth]{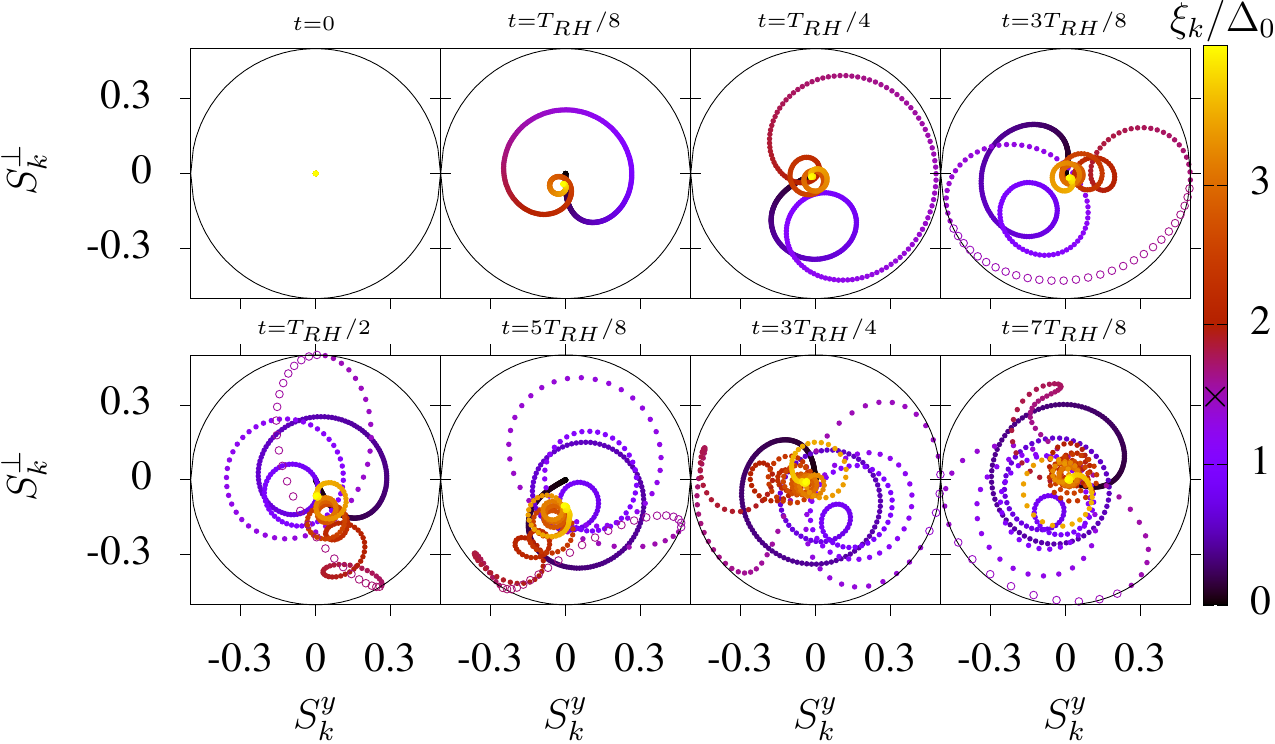}
\caption{(Color online) Dynamics in the Rabi-Higgs phase. We show snapshots, during the first Rabi period $T_{RH}$, of pseudospins dynamics in the Rabi-Higgs regime ($\alpha=0.12$ and $\omega_d=3.5\Delta_0$ corresponding with the parameters used in Fig.~\ref{fig:lambdadyn} (e,f)). On the top of panels, we indicate the time. Those pseudospins with $S^z_k>0$ ($S^z_k<0$) are indicated with filled (empty) dots. The cross in the colour bar indicates the $\xi^*_{\bm k}$ satisfying the resonance condition $\omega_L(\xi_{\bm k}^{\ast})=\omega_d$. See also Supplementary Video~3.}
\label{fig:snapshotRH}
\end{figure*}

\subsection{Discrete time-crystal phase}
We now turn to the discrete time-crystal phase. Because $\bar{\Delta}=0$ is very far from the equilibrium value $\Delta_0$, it is convenient to use a Cartesian frame instead of the ELF. 
Figure~\ref{fig:snapshotTC} shows the dynamics  during a 2$T_d$ time window.
Full (open) symbols indicate $S^z_{\bm k}>0$  ($S^z_{\bm k}<0$).
The high-energy pseudospins (red-yellow dots) precess around the instantaneous pseudomagnetic field describing a loop (green line) and contributing self-consistently to build the time-dependent order parameter. Very low energy pseudospins have a nearly maximal component in the $xy$ plane in the first frame, indicating strong pairing correlations but with incoherent phases. This initial circular feature becomes an ellipse at subsequent times, corresponding to a ring that rotates nearly rigidly along the $x$ axis of the Bloch sphere. Spins at intermediate energies (violet-orange dots) interpolate between these two behaviors, contributing significantly to the time dependent $\Delta(t)$.
Indeed, the red and violet cloud is on the right side of the frame at $t=0$ contributing to a positive $\Delta$ [cf. Eq.~(\ref{eq:deltat})] and 
after one drive period has shifted to the left, yielding the sign alternation of $\Delta$ in one drive period as shown in Fig.~\ref{fig:phasedDOS}(c).
After two drive periods, the pattern goes approximately back to the original distribution, consistently with the behavior of the order parameter.  

\subsection{Rabi-Higgs phase}
Turning now to the Rabi-Higgs phase, since the order parameter is close to equilibrium, it is convenient to use the ELF once again. In Fig.~\ref{fig:snapshotRH} 
we show several texture snapshots during the first Rabi-Higgs period $T_{RH}$. At $t=0$ (equilibrium) the pseudospins texture corresponds to all pseudospins at the origin by definition. For $t=T_{RH}/2$ we have the inversion phenomenon in which a subset of pseudospins (empty dots) get inverted with respect to its fermionic energy. In other words, they have an instantaneous negative $S^z_{\bm{k}}$  while $\xi_{\bm{k}}>0$. Since the $z$ component of the pseudospin encodes
the charge, this corresponds to an inversion of the quasiparticle population.
    The pseudospins that get inverted satisfy the resonance condition $\omega_d=\omega_L (\xi_{\bm{k}}^{\ast})$ with $\omega_L (\xi_{\bm{k}})=2\sqrt{\xi_{\bm{k}}^{2}+\Delta_0^{2}}$, the natural Larmor frequency. We indicate this fermionic energy $\xi_{\bm{k}}^{\ast}$ with a cross in the colour bar. After one full Rabi period, this inversion gets largely diminished. 
In the steady state one observes a periodic oscillation of the population as discussed in Ref.~\cite{OjedaCollado2018}. 

\subsection{    Gapless phase}
\begin{figure*}[tb]
\includegraphics[width=0.8\textwidth]{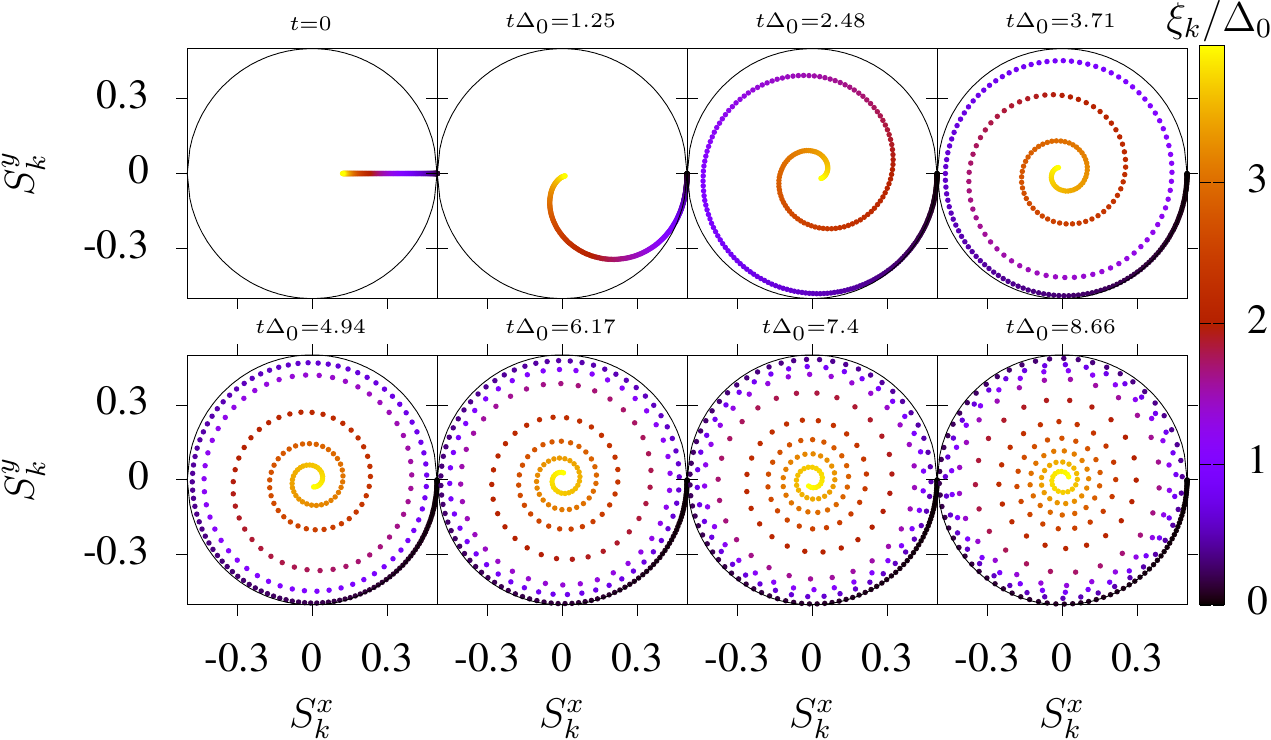}
\caption{(Color online) Dynamics in the gapless regime. We show snapshots, during the transient to the gapless behavior, of pseudospins dynamics in the gapless phase ($\alpha=0.27$ and $\omega_d=3.5\Delta_0$ corresponding with the parameters used in Fig.~\ref{fig:lambdadyn} (g,h)). On the top of panels, we indicate the time. See also Supplementary Video~4.}
\label{fig:snapshotGL}
\end{figure*}

Finally, for the gapless phase it is convenient to go back to the Cartesian frame. In Fig.~\ref{fig:snapshotGL} we show texture snapshots at short times. 
We see  that after a fast transient, the pseudospins start to roll up around the coordinate origin with the shape of a spiral. This unveils that the gapless phase consists in strong pairing correlations, i.e. for several pseudospins  $S_{\bm{k}}^{x}\neq 0$, but with Cooper pairs which are not phase-coherent. So the sum of the $x$ components yields a zero gap. This is not, however, a chaotic state, but the ZOPA 
is a consequence of a very orderly movement of pseudospins consistent with the unitary evolution of the state.

%%%%%%%%%%%%%%%%%%%%%%%%%%%%%% 
\section{Dynamical Phase Transitions}\label{sec:dynam-phase-trans}
\begin{figure}[tb]\label{fig:dpt}
\includegraphics[width=0.5\textwidth]{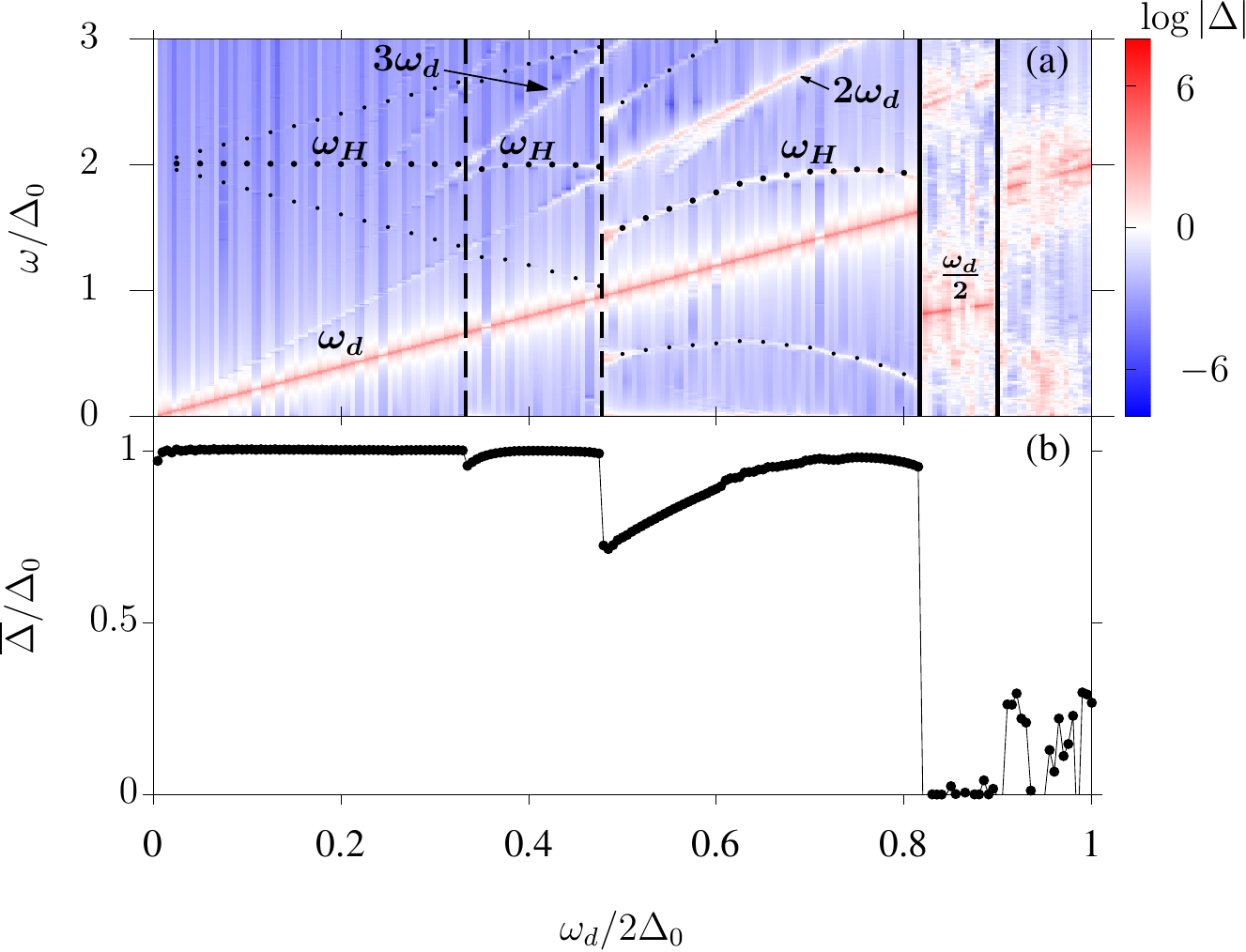}
\caption{(Color online) (a) Characterization of DPTs thought a false colour plot of the superconducting response in the frequency domain vs. driving frequency,  $\omega_d$ for $\alpha=0.05$.
We used $t\Delta_0 \in [1500, 2500]$ for high frequency resolution.
  Large and small dots represent $2\bar{\Delta}$ (shown also in b) and $2\bar{\Delta}\pm\omega_d$ respectively. These values are taken from  
the real time gap dynamics $\Delta(t)$. Notice that they match peaks in the FT. Vertical lines indicate equal symmetry (dashed) and different symmetry (full) first order DPTs. (b) Average of superconducting order parameter vs. drive frequency. } 
\label{fig:verticalcut}
\end{figure}

%%%%%%%%%%%%%%%%%%%%%%%%%%%%%%
In order to characterize the order of the different DPTs, we analyse the behaviour of the system along the dashed vertical line  in Fig.~\ref{fig:phased}~(a)
corresponding to $\alpha=0.05$. Figure~\ref{fig:verticalcut} (a) shows the FT of the time-dependent superconducting order parameter as a function of $\omega_d$ while panel (b) shows its long term average in the steady state, $\bar{\Delta}$.

In the adiabatic limit ($\omega_d\ll 2\Delta_0$),  the FT shows a clear peak at $\omega_d$ resulting in the strong linear orange feature in Figure~\ref{fig:verticalcut}(a) as expected from linear response. A weaker feature appears at $2\omega_d$ corresponding to the allowed~\cite{OjedaCollado2020} second harmonic generation.
Increasing $\omega_d$, the dynamical order parameter  $\bar{\Delta}$ [panel (b)] shows discontinuities near $\omega_d/(2\Delta_0)=1/2$ and $\omega_d/(2\Delta_0)=1/3$
corresponding to first order DPTs associated to low $\alpha$ precursors of the Arnold tongues [c.f. Fig.~\ref{fig:phased} (a)].  
Inside the dynamical phases with finite $\bar{\Delta}$, well-defined peaks appear at $\omega_H$ and $\omega_H\pm\omega_d$ associated with the synchronized Higgs mode.  The dotted lines in panel (a) are $2\bar{\Delta}$ (large dots) and $2\bar{\Delta}\pm \omega_d$ (small dots), showing that the frequency of the Higgs mode is locked at $2\bar \Delta$. This can be seen as  incommensurate time-crystal behavior similar to Ref.~\cite{Homann2020}.

For $2\omega_d$  higher than the minimum of the effective continuum $2\bar{\Delta}$, the second harmonic of the drive is resonant with the quasiparticles. This produces a  proliferation of excitations resulting in a suppression of the average gap as shown in Fig.~\ref{fig:verticalcut} (b) for $\omega_d/2 \Delta_0\approx1/2$. As $\omega_d$  decreases, 
the $2\omega_d$ resonance approaches the quasiparticle minimum and the range of resonant quasiparticles gets cutoff. At some point, there are not enough quasiparticles with $2\sqrt{\xi_{\mathbf{k}}^2+\bar{\Delta}^2}\approx 2\omega_d$ to suppress the gap and a new steady state is found, in which the Higgs mode frequency increases discontinuously with $\omega_H>2\omega_d$. Thus, the line $\omega=2\omega_d$ intersects the jump of $\omega_H$ in Fig.~\ref{fig:verticalcut} (a). 
A similar mechanism applies to the weaker transition at lower frequency, with
the third harmonic resonance and the line $\omega=3\omega_d$ intercepting the jump of $\omega_H$. 

\begin{figure}[tb]
\includegraphics[width=0.5\textwidth]{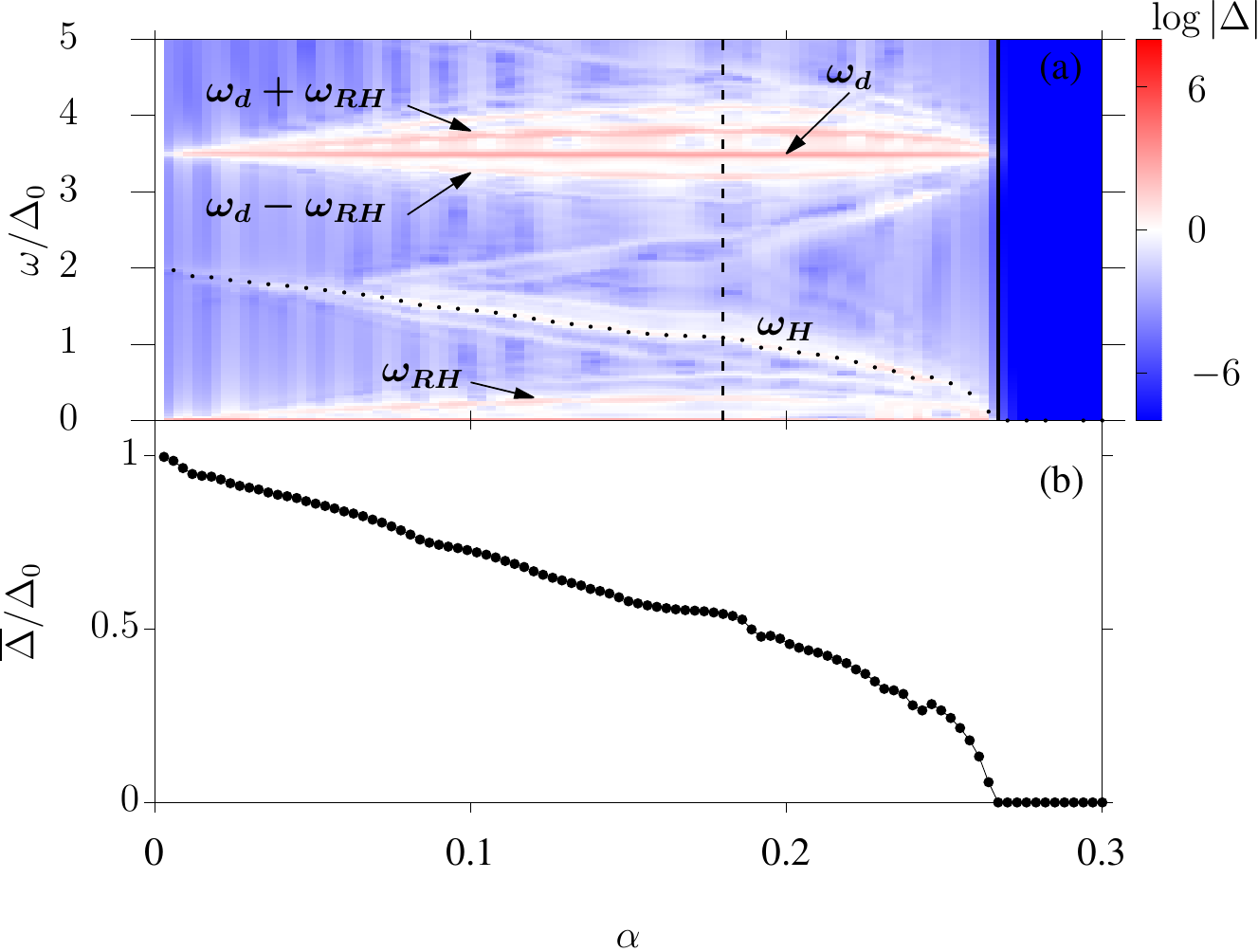}
\caption{(Color online) (a) The superconducting response in the frequency domain vs. the amplitude of the drive $\alpha$ for $\omega_d=3.5\Delta_0$.
We used $t\Delta_0 \in [10, 200]$ to compute the average and the FT. Small dots represent $2\bar{\Delta}$ (shown also in b). These values are taken from the real time gap dynamics $\Delta(t)$ and match peaks in the FT. Dashed vertical line indicates the $\alpha$ value for which the Rabi-Higgs mode start to soften. Full vertical line instead points out the second order DPTs to a gapless regime. The blue region on the right corresponds to $\Delta=0$. (b) Average of superconducting order parameter vs. drive amplitude.} 
\label{fig:horizontalcut}
\end{figure}

While the previous two DPTs are among phases with the same symmetry, 
the third discontinuity at $\omega_d/2\Delta_0\simeq 0.8$ represents a DPT between qualitatively distinct phases:  gaped on the left and ZOPA
on the right. The ZOPA phase corresponds to the $n=1$  Arnold tongue and is bounded by the solid vertical lines in Fig.~\ref{fig:horizontalcut} (a).
Inside this region ($0.8 \lesssim \omega_d/2\Delta_0 \lesssim 0.9$), a commensurate time-crystal appears with a doubling of the period of the drive.  
The persistence of the order parameter oscillation at $\omega_d/2$ in this finite region indicates that the period doubling is not accidental. As discussed in Ref.~\cite{OjedaCollado2021} this is also true changing $\alpha$ 
in a finite range. Thus, we confirm again that this state  satisfies the rigidity criteria~\cite{Yao2017} for time-crystalline behaviour.

Both transitions bounding the time-crystal phase are first order. 
The upper edge of the tongue has a fractal like appearance~\cite{Collado2021} which has been noticed also in related models~\cite{Giachetti2022}.  Even inside the tongue, as already mentioned, marbled textures appear where $\bar{\Delta}$ becomes nonzero. In these cases, as illustrated in the rightmost part of  Fig.~\ref{fig:verticalcut} (a), the time crystal is lost and the system switches to a phase in which time-translational symmetry is recovered at $\omega_d/2\Delta_0\sim 0.9$ (marked with a second vertical line in the figure). For higher frequencies, the system enters a complex regime in which our numerical calculations show instabilities. In general, we find that when $\bar{\Delta}\simeq 0$ (for $\omega_d<2\Delta_0$) a time-crystal phase emerges. 

%%%%%%%%%%%%%%%%%%%%%%%%%%%%%%%%%%%%%%%%%%%%%%%%%%%%%
%%%%%%%%%%%%%%%%%%%%%%%%%%%%%%%%%%%%%%%%%%%%%%%%%%%%%%%%

Now we discuss on the DPTs along the horizontal dashed line $\omega_d=3.5\Delta_0$ shown in Fig.~\ref{fig:phased} (a).  Figure~\ref{fig:horizontalcut} (a) shows the FT map of $\Delta(t)$ for different values of the drive amplitude $\alpha$. For extremely weak perturbation amplitudes, the system essentially responds with the drive frequency and the Higgs mode $\omega_H=2\bar{\Delta}$ as occurs for drive frequencies below the gap. As soon as we increase the drive amplitude $\alpha$, the Rabi-Higgs mode $\omega_{RH}$ appears in the spectrum, whose frequency  increases linearly by increasing $\alpha$. At the same time,  satellites at frequencies $\omega_d\pm n \omega_{RH}$ with $n=1,2$ appear in the FT. Slightly less prominent, but still appreciable, are peaks at $\omega_H\pm\omega_{RH}$  witnessed by white branches near to $\omega_H$ that linearly grow with $\alpha$.

Increasing even more the  amplitude, it is clear from Fig~\ref{fig:horizontalcut} (a) that there is a critical value $\alpha$ (marked by the vertical dashed line) in which the Rabi-Higgs starts to soften. It is more clearly seen in the satellites peaks $\omega_d\pm\omega_{RH}$. This is an {\em anomalous} behaviour, taking into account that a conventional Rabi frequency increases by increasing the amplitude of the drive. Increasing
even more the drive, the Rabi-Higgs and Higgs modes disappear and the system enters into a gapless regime (blue area in panel a). In this case, not only $\bar{\Delta}=0$ but also $\Delta=0$ (i.e. the superconducting order parameter is zero over time without exhibiting oscillations). This DPT is indicated by a solid vertical line. In contrast to the  first order DPTs described in Fig.~\ref{fig:verticalcut}, here both the dynamics and the order parameter $\bar{\Delta}$, shown in panel (b), point to a second order DPT with the characteristic ``critical slowing down'' near the transition point. 

%describing our different dynamical phases, we realize that in contrast with the previous first order DPTs described in Fig.~\ref{fig:verticalcut}, here $\bar{\Delta}$ does not show discontinuities but its derivate. In this case we have a second order DPT from Rabi-Higgs to gapless phase. 

%%%%%%%%%%%%%%%%%%%%%%%%%%%%%%%%%%%%%%%%%%%%%%%%%%%%%
%%%%%%%%%%%%%%%%%%%%%%%%%%%%%%%%%%%%%%%%%%%%%%%%%%%%%%%%

%%%%%%%%%%%%%%%%%%%%%%%%%%%%%%%%%%%%%%%%5
\section{Mapping to classical dynamics}
\label{sec:mapclas}
%%%%%%%%%%%%%%%%%%%%%%%%%%%%%%%%%%%%%%%%%%%%%%%%%%%%%%%%%%%%%%%%%%%%%%
In order to investigate the origin of the parametric resonances found numerically in Fig.~\ref{fig:phased} it is useful to map the  BCS dynamics to classical anharmonic oscillators~\cite{Sciolla2011}.  This can be done because the dynamics of pseudospins can be mapped to the dynamics of a collection of \emph{classical} spins $\bm{S}_{\bm k}$. Such dynamic is governed by the Hamilton equations $\partial_{t}\bm{S}_{{\bm k}}=\left\{ \mathcal{H},\bm{S}_{{\bm k}}\right\}$  using the usual Poisson brackets $\left\{ S_{{\bm k}}^{\mu},S_{{\bm k}'}^{\nu}\right\} =-\varepsilon_{\mu\nu\eta}\delta_{{\bm k},{\bm k}'}S_{{\bm k}}^{\eta}$ for angular momenta where $\mu, \nu, \eta$ represents the $x, y, z$ component and $\varepsilon_{\mu\nu\eta}$ is the Levi-Civita tensor.

%\begin{equation}
 % \mathcal{H}=-\sum_{{\bm k}}2\xi_{{\bm k}}S_{{\bm k}}^{z}-\lambda\sum_{{\bm k},{\bm k}'}S_{{\bm k}}^{+}S_{{\bm k}{'}}^{-}-2f \sum_{\bm k} S^x_{\bm k}.
%\end{equation}

Alternatively one can derive the BCS time-dependent equations from a variational principle, requiring that the wave-function has a BCS form at each instant of time and using the elements of the generalized one-particle density matrix as dynamical variables ~\cite{Blaizot1986,Schiro2010,Sciolla2011,Bunemann2013,Seibold2021}. For each pair ($\bm{k}\uparrow$, $-\bm{k}\downarrow$), four expectation values of the one-particle density matrix need to be considered, corresponding to the four operators appearing on the right-hand side of Eqs.~\eqref{eq:pesudospins}.

One can show that the density matrix derives from a BCS state if and only if the generalized matrix is idempotent~\cite{Blaizot1986}. It is easy to show that this is equivalent to the following constraints, 
\begin{eqnarray}
      \label{eq:cons1}
&&\left(S_{{\bm k}}^{x}\right)^{2}+\left(S_{{\bm k}}^{y}\right)^{2}+\left(S_{{\bm k}}^{z}\right)^{2}=1/4,\\ &&\langle\hat{c}_{\bm{k}\uparrow}^{\dagger}\hat{c}_{\bm{k}\uparrow}^{}\rangle=\langle\hat{c}_{-\bm{k}\downarrow}^{\dagger}\hat{c}_{-\bm{k}\downarrow}^{}\rangle.\label{eq:cons2}
\end{eqnarray}
We can use Eq.~\eqref{eq:cons2} to reduce the dynamical variables for a pair ($\bm{k}\uparrow$, $-\bm{k}\downarrow$) to three variables which can then be taken as the pseudospin expectation values $S_{{\bm k}}^{x}$, $S_{{\bm k}}^{y}$, $S_{{\bm k}}^{z}$ with the constraint Eq.~\eqref{eq:cons1}.

%From the equations of motion (EOM) one can show that $S_{{\bm k}}^{x}\dot{S_{{\bm k}}^{x}}+S_{{\bm k}}^{y}\dot{S_{{\bm k}}^{y}}+S_{{\bm k}}^{z}\dot{S_{{\bm k}}^{z}}=0$ implying that the length of spins,
%are conserved quantities. 

In this formalism, the time-dependent expectation value of the quantum Hamiltonian plays the role of a 
 classical Hamiltonian and can be obtained from Eq.~(\ref{eq:ham}) replacing operators by their expectation values in the instantanous BCS wave-function.  Adding the constraint Eq.~(\ref{eq:cons1}) with Lagrange multipliers $\omega_{\bm k}$, the classical Hamiltonian reads,
\begin{eqnarray}
  \label{eq:de}
\mathcal{H}&=& - \sum_{\bm k} 2 S^z_{\bm k} \xi_{\bm k} -\lambda \sum_{{\bm k},{\bm k}'} (S^x_{\bm k} S^x_{{\bm k}'}+S^y_{\bm k} S^y_{{\bm k}'})\\
&-&2f \sum_{\bm k} S^x_{\bm k}+\sum_{\bm k} \omega_{\bm k} \left[(S^x_{\bm k})^2+(S^y_{\bm k})^2+(S^z_{\bm k})^2-\frac14\right]\nonumber
\end{eqnarray}
The third term is an additional external pairing field which couples linearly with the order parameter and which we added for latter use. As before, driving will be introduced by making  $\xi_{{\bm k}}$, $\lambda$, or $f$, time-dependent.  Here we concentrate on the $\lambda$-driving and discuss briefly $f$-driving. The DOS-driving protocol can be treated similarly.  

It is useful to use as dynamical variables the deviation (not necessarily small)
$\delta S^\mu_{\bm k}$ defined as $S^\mu_{\bm k}\equiv S^{\mu,0}_{\bm k}+\delta S^\mu_{\bm k}$, with $S^{\mu,0}_{\bm k}$ the equilibrium BCS state.  We also write the time dependence of the pairing interaction
as an average value  $\lambda_0$ plus  a fluctuation, $\lambda=\lambda_0+\delta \lambda$. The Hamiltonian can be written as $\mathcal{H}=E_0+\delta\mathcal{H}$
with $E_0$ the equilibrium BCS ground state energy and 
with the fluctuating part, 
\begin{eqnarray}
  \label{eq:fluc}
\delta \mathcal{H}=& -&2 \sum_{\bm k} \delta S^z_{\bm k} \xi_{\bm k} - \delta\lambda \sum_{{\bm k},{\bm k}'} (S^{x,0}_{\bm k} S^{x,0}_{{\bm k}'}+S^{y,0}_{\bm k} S^{y,0}_{{\bm k}'})\nonumber\\
&-&2(\lambda_0+\delta\lambda) \sum_{{\bm k},{\bm k}'} (\delta S^x_{\bm k} S^{x,0}_{{\bm k}'} + \delta S^y_{\bm k} S^{y,0}_{{\bm k}'})  \nonumber\\
&-& (\lambda_0+ \delta\lambda) \sum_{{\bm k},{\bm k}'} (\delta S^x_{\bm k}\delta S^x_{{\bm k}'}+\delta S^y_{\bm k}\delta  S^y_{{\bm k}'})\\
&+&\sum_{\bm k} \omega_{\bm k} [2S^{x,0}_{\bm k} \delta S^x_{\bm k}+2S^{y,0}_{\bm k} \delta S^y_{\bm k}+2S^{z,0}\delta S^z_{\bm k}\nonumber\\
&+&  (\delta S^x_{\bm k})^2+(\delta S^y_{\bm k})^2+(\delta S^z_{\bm k})^2]-2f \sum_{{\bm k}} \delta S^{x}_{\bm k} \nonumber.
\end{eqnarray}

The saddle point condition requires that linear-variations vanish which,
by setting $f=\delta \lambda=0$, yields
the equilibrium mean-field equations, 
\begin{eqnarray}
  \label{eq:mf}
  -\Delta^x_0 +\omega_{\bm k} S^{x,0}_{\bm k}&=&0,\\
  -\Delta^y_0 +\omega_{\bm k} S^{y,0}_{\bm k}&=&0,\\
  -\xi_{\bm k} +\omega_{\bm k} S^{z,0}_{\bm k}&=&0,
\end{eqnarray}
with $\Delta^\mu_0=\lambda_0 \sum_{\bm k} S^{\mu,0}_{\bm k}$ which can be readily solved for 
$S^{\mu,0}_{\bm k}$. Without loss of generality, we take $\Delta_0^x=\Delta_0$ and
 $\Delta_0^y=0$. Applying the constraint to the stationary state, one finds that the Lagrange multiplier is given by the equilibrium Larmor frequency, 
 $\omega_{\bm k}=\omega_{L}(\xi_{\bm{k}})=2\sqrt{\xi^2_{\bm k}+\Delta_0^2}$. The negative root can be discarded as it yields the unphysical sign of $S^{z,0}_{\bm k}$. Solving for the spin components, yields Eq.~(\ref{eq:sequilib}).

Using the saddle point condition, the Hamiltonian has terms up to cubic in fluctuations  and is given by
\begin{eqnarray}
\label{eq:fluc2}
\delta \mathcal{H}=&-& (\lambda_0+ \delta\lambda) \sum_{{\bm k},{\bm k}'} (\delta S^x_{\bm k}\delta S^x_{{\bm k}'}+\delta S^y_{\bm k}\delta  S^y_{{\bm k}'})\nonumber\\
                   &+&\sum_{\bm k} \omega_{\bm k} [(\delta S^x_{\bm k})^2+(\delta S^y_{\bm k})^2+(\delta S^z_{\bm k})^2]\\
&-&2\left(\frac{\delta\lambda}{\lambda_0}\Delta_0+f\right) \sum_{{\bm k}} \delta S^x_{\bm k} \nonumber,
\end{eqnarray}
where we droped terms linear in $\delta \lambda$ which do not affect the EOM. It is useful to transform the Hamiltonian to the  ELF of Eq.~(\ref{eq:defs}) to obtain,  
\begin{eqnarray}
  \label{eq:bcsfluc}
\delta \mathcal{H}&=&  \sum_{\bm k} \omega_{\bm k} [(\delta S^y_{\bm k})^2+(\delta S^\perp_{\bm k})^2+ (\delta S^\parallel)^2]\nonumber\\
&-& (\lambda_0+\delta\lambda) \sum_{{\bm k},{\bm k}'}\left(4\frac{\xi_{\bm k}\xi_{{\bm k}'}}{\omega_{\bm k}\omega_{{\bm k}'}}
\delta S^\perp_{\bm k}\delta S^\perp_{{\bm k}'}+\delta S^y_{\bm k}\delta  S^y_{{\bm k}'}\right)\nonumber\\
&-&4\left(\frac{\delta\lambda}{\lambda_0}\Delta_0+f\right) \sum_{{\bm k}} \frac{\xi_{\bm k}}{\omega_{\bm k}} \delta S^\perp_{\bm k}.
\end{eqnarray}

\subsection{Harmonic approximation}
So far, the treatment is exact. 
We now proceed by introducing some approximations. First, one can use the constraint to show that longitudinal fluctuations are higher order compared to transverse ones
(which can be also seen from a geometric argument). Thus,  we neglect $\delta S_{\bm k}^\parallel$ in the first term of Eq.~(\ref{eq:bcsfluc}).
Defining canonical variables as 
$p_{\bm k}=\sqrt2 \delta S^y_{\bm k}$ and $q_{\bm k}=\sqrt2 \delta S^\perp_{\bm k}$
the energy reads, 
\begin{eqnarray}
  \label{eq:bcsfluccano}
\delta \mathcal{H}&=& \frac12 \sum_{\bm k} \omega_{\bm k} (p_{\bm k}^2+q_{\bm k}^2)\nonumber\\
&-& (\lambda_0+\delta\lambda) \sum_{{\bm k},{\bm k}'}\left(2\frac{\xi_{\bm k}\xi_{{\bm k}'}}{\omega_{\bm k}\omega_{{\bm k}'}}
q_{\bm k} q_{{\bm k}'}+ \frac12  p_{\bm k}p_{{\bm k}'}\right) \nonumber\\
&-&2\sqrt2
\left(\frac{\delta\lambda}{\lambda_0}\Delta_0+f\right)\sum_{{\bm k}}     \frac{\xi_{\bm k}}{\omega_{\bm k}} q_{\bm k}
\end{eqnarray}
which maps the problem to a set of harmonic oscillators with long-range interactions depending on the quasiparticle energy $\xi_{\bm k}$.
The Larmor frequency $\omega_{L}(\xi_{\bm{k}})$ plays the role of natural frequencies of the oscillators, giving rise to a DOS in frequency space peaking at $2\Delta_0$,  consistent with the identification of $\omega_0=2\Delta_0$ as the ``natural frequency'' of oscillation of the superconducting system~\cite{OjedaCollado2021}.

Setting $f=0$, one can check that the Hamilton EOM,
\begin{equation}
  \label{eq:heom}
  \dot{q}_{\bm k}=\frac{\partial\mathcal{ H}}{\partial p_{\bm k}},\ \ \ \ \ \ 
   \dot{p}_{\bm k}=-\frac{\partial\mathcal{ H}}{\partial q_{\bm k}}   
\end{equation}
reproduce the linearized version of Eq.~\ref{eq:eom}.% in the main text. 

%%%%%%%%%%%%%%%%%%%%%%%%%%%%%%%%%%%%%%%%%%%%%%%%%%%%%
\begin{figure}[tb]
\includegraphics[width=0.5\textwidth]{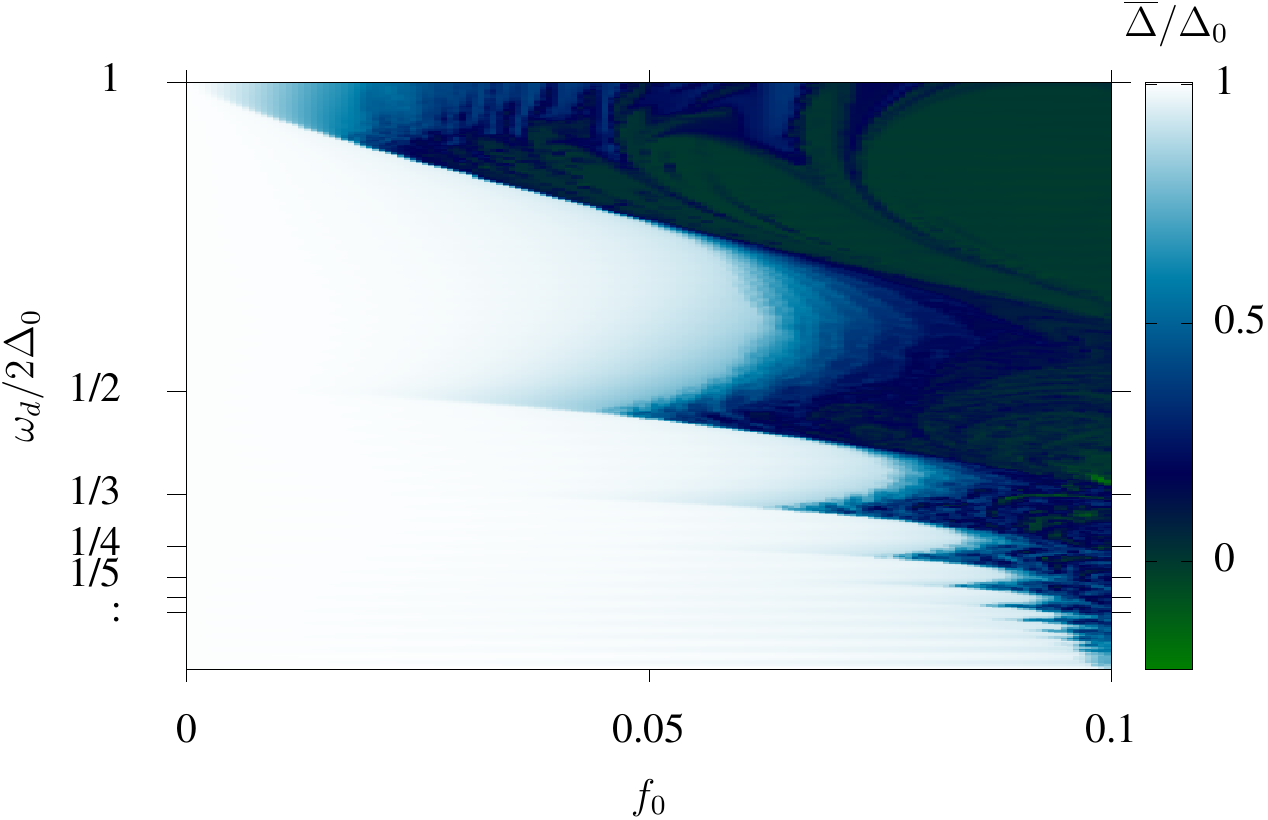}
\caption{(Color online)  Dynamical phase diagram for periodic driving by an external pairing field. We show the temporal average of the superconducting order parameter in the 
  $f_0$-$\omega_{d}$ plane for the case of an external paring field drive.  }
\label{fig:figs1}
\end{figure}
%%%%%%%%%%%%%%%%%%%%%%%%%%%%%%%%%%%%%%%%%%%%%%%%%%%%%%%%

Fluctuations in $\lambda$ couple linearly with the canonical variables [last term in Eq.~(\ref{eq:bcsfluccano})] and with quadratic fluctuations of canonical variables (second term).   
In analogy with a single classical parametric oscillator~\cite{Landau1976}, it is tempting to attribute parametric resonances to the coupling with quadratic fluctuations. However, this can be excluded in the following way. We set $\delta\lambda=0$ and consider periodic driving in 
$f=f_0 \sin(\omega_d t)$. From Eq.~\eqref{eq:bcsfluccano}, we see that this driving is equivalent to the linear coupling with $\delta\lambda$ (for that term). Fig.~\ref{fig:figs1} shows the
phase diagram  computed with the \textit{original} Hamiltonian Eq.~(\ref{eq:de})
but with $f$ as the only time-dependent perturbation. 
In this case, one obtains similar Arnold tongues which reveals that the coupling of $\delta\lambda$ with quadratic fluctuations is not essential to obtain the parametric resonances.  
This result could be anticipated from the following argument. 
Since  quadratic fluctuations of canonical variables couple linearly with $\delta\lambda$ [c.f. Eq.~(\ref{eq:bcsfluccano})], the 
analogy with a single parametric oscillator would yield 
parametric resonances at  $\omega_d=2\omega_0/n=4\Delta_0/n$ which is not consistent with the resonances obtained numerically [Fig.~\ref{fig:phased} (a)] which satisfy instead,   $\omega_d=\omega_0/n=2\Delta_0/n$. 

We will see that the linear coupling of $\delta\lambda$ with canonical variables [last term in Eq.~(\ref{eq:bcsfluccano})] plays a fundamental role in the emergence of the parametric resonances. This does not occur through the direct coupling with the pseudospins fluctuations but indirectly through the effect of higher order non-linearities.

\subsection{Higher orders}
Writing the constraint on the pseudospins lenght as $2{\bm S}_{\bm k}^0.\delta {\bf S}_{\bm k}+ |\delta {\bf S}_{\bm k}|^2=\delta S_{\bm k}^\parallel+|\delta {\bf S}_{\bm k}|^2=0$, we can eliminate longitudinal fluctuations in favour of transverse ones,
\begin{equation}
  \label{eq:root}
\frac12+\delta S_{\bm k}^\parallel=\pm \frac12  \sqrt{1-4(\delta S^y_{\bm k})^2+(\delta S^\perp_{\bm k})^2},  
\end{equation}
where the left-hand side is nothing but $S_{\bm k}^\parallel$. We see that for the solution to be real $(\delta S^y_{\bm k})^2+(\delta S^\perp_{\bm k})^2<1/4$ and two solutions are possible, one in which  $S_{\bm k}^\parallel>0$ (ferro alignment) and one in which $S_{\bm k}^\parallel<0$
antiferro alignment. %We know that during the non-linear dynamics both solutions are relevant. For
Although both solutions are needed for the full dynamics, 
for the time being we restrict to ferro-alignment which is the relevant solution for not too large deviations.

Considering only the $\lambda$-driving and using canonical variables $q_{\bm k}$ and $p_{\bm k}$, the Hamiltonian reads, 
\begin{eqnarray}
  \label{eq:bcsflucnonlincano}
\delta \mathcal{H}&=&  \sum_{\bm k}  \frac{\omega_{\bm k}}2 \left(1-\sqrt{1-2(p_{\bm k}^2+q_{\bm k}^2)}\right)\nonumber\\
&-& (\lambda_0+\delta\lambda) \sum_{{\bm k},{\bm k}'}\left(2\frac{\xi_{\bm k}\xi_{{\bm k}'}}{\omega_{\bm k}\omega_{{\bm k}'}}
q_{\bm k} q_{{\bm k}'}+ \frac12  p_{\bm k}p_{{\bm k}'}\right)
\nonumber\\
&-&2\sqrt2\frac{\delta\lambda}{\lambda_0}\Delta_0 \sum_{\bm k}     \frac{\xi_{\bm k}}{\omega_{\bm k}} q_{\bm k}.
\end{eqnarray}
where again one can check that Hamilton equations yield the correct EOM. 

Expanding the square root in a Taylor series, the first term in Eq.~(\ref{eq:bcsflucnonlincano}) can be written as a collection of anharmonic oscillators, 
%\begin{equation}
$\sum_{\bm k}\frac{\Omega_{\bm k}}{2}(p_{\bm k}^{2}+q_{\bm k}^{2}),$
%\end{equation}
with a time-dependent self-consistently determined natural frequency, 
\begin{eqnarray}
  \label{eq:freqmodule}
\Omega_{\bm k}(t)=\omega_{\bm k}\left[\right.1+\frac{1}{2}(p_{\bm k}^{2}+q_{\bm k}^{2})+\frac{1}{2}(p_{\bm k}^{2}+q_{\bm k}^{2})^{2}
\nonumber\\
+\frac{5}{8}(p_{\bm k}^{2}+q_{\bm k}^{2})^{3}+\ldots\left.\right].
\end{eqnarray}
Following Ref.~[\onlinecite{Landau1976}] we can analyse nonlinearities iteratively.
Treating the last term in Eq.~(\ref{eq:bcsflucnonlincano}) in linear response~\cite{OjedaCollado2018} one obtains that for a drive frequency $\omega_d$, the quadratic terms respond as
$(p_{\bm k}^{2}+q_{\bm k}^{2})\propto \cos{2\omega_d t}$ which implies that the natural frequency $\Omega_{\bm k}$ is ``pumped'' with frequency $\omega_p=2\omega_d$. Thus, using the fact that a classical oscillator has parametric resonances at $\omega_d=2\omega_0/n$, one expects resonances at $\omega_d=2\Delta_0/n$ in the BCS system, as indeed found. The identification of the pump-mechanism and the explanation of the factor of two in the resonance series is the main result of this section.  We remark that this ingredient alone is not enough to explain the parametric resonances. Indeed, the non-linear effects described arise from the constraint while in  Sec.~\ref{sec:import-inter} we showed that interactions, represented here by the second line in Eq.~(\ref{eq:bcsflucnonlincano}) are essential to stabilize parametric resonances. Furthermore,  restricting to the ``ferro'' alignment root in Eq.~(\ref{eq:bcsflucnonlincano}) is not enough in the parametric resonance regime as numerically we find that the dynamics explore both roots [see Fig.~\ref{fig:snapshotTC}]. On the other hand, 
these considerations, do not affect the conclusion  that the leading pump frequency for the oscillators is at $\omega_p=2\omega_d$ instead of $\omega_d$ as adding more non-linearites can only produce higher multiples of $\omega_d$.
%as there is no way to generate this linear coupling adding higher order non-linearities.

\section{Summary and Conclusions}
\label{sec:conclusions}

We have presented a comprehensive dynamical phase diagram for a periodically driven BCS condensate using different driving protocols.
We concentrated on superconducting/superfluid phases but our results are valid for any phase for which a BCS description is valid in the
time regime before energy relaxation process take place in the system.
This includes weak-coupling spin and charge density waves which can also be mapped to the BCS model.

We numerically demonstrated that the existence of four dynamical phases and parametric resonances are quite robust to changes in the  protocol.
To a large extent, the phase diagram can be said universal. We expect the main features to remain, also, for more complicated drives, which for example may be anisotropic on the Fermi surface and depend on light polarization acting as drive.

A detailed analysis of the evolution of the pseudospin textures (Figs.~\ref{fig:snapshotHiggs}-\ref{fig:snapshotGL}) allowed to visualize how, the many-body system spontaneously self-organize in momentum space in sectors with different dynamics. For the gapless phase, our analysis revealed a remarkable degree of order and symmetry in this, apparently, unbroken symmetry phase. 

Our study allowed to  identify the order of the phase transitions. Roughly speaking,  parametric resonances at 
$2\Delta_0/n$ can be seen as multiphoton process with $n$ photons reaching the gap. Continuous excitation produces a depletion of the gap which lowers the threshold for excitation. This provides a feedback loop that explains the first order transitions in the lower edge of the Arnold tongues. Second order phase transitions also arise by increasing the intensity of the drive from the Rabi-Higgs to the gapless phase. This is accompanied by a critical slowing down of the dynamics. Thus, the Rabi-Higgs frequency first follows the common expectation for a Rabi mode; its frequency increases with dive strength. For large driving strength it switches to an anomalous regime in which its frequency decreases with strength.

For subgap excitations, by combining different driving protocols and mapping to a system of non-linear oscillators, we showed that the mechanism of parametric pump can be traced back to the non-linearity in the system due to the constraint on the length of pseudospins. Furthermore, we demonstrated that to take into account interactions self-consistently is an essential ingredient to obtain parametric resonances. In other words, they constitute an emergent phenomenon of the many-body system.

Regarding experimental realizations, while the present model neglects the presence of a bath, we have previously shown that, parametric resonances and time-crystal behaviour, appear at very early times in the dynamics~\cite{OjedaCollado2021}. In the presence of a finite decoherence time, they may be observed as a prethermal transient in the dynamics. Proposed realizations in the solid state include THz excitations and a phonon assisted $\lambda$ driving mechanism~\cite{OjedaCollado2018}.
Ultracold atoms are also very interesting platforms which are inherently much less affected by the environment (lattice vibrations are not present). Furthermore, they offer a large degree of parameter manipulation, making them ideal candidates to study the driven BCS system~\cite{Behrle2018}.
Yet another promising platform to observe the present phenomena is a cavity-QED
simulator, where Anderson's pseudospins model can be directly studied~\cite{Lewis-Swan2021}.
Our work is an invitation to exploit these platforms to experimentally explore the fascinating time-translational symmetry breaking phases of periodically driven BCS systems.

%and quantum simulators~\cite{Lewis-Swan2021}. We have analyzed in detail the problem of Floquet engineering in BCS systems and our results are of interest not only for nonequilibrium superconductivity but spin-density waves and charge density waves in the weak coupling regime at short times before the energy relaxation process take place in the system.

%%%%%%%%%%%%%%%%%%%%%%%%%%%%%%%%%%%%%%%%%%%%%%
\begin{acknowledgments}
We acknowledge financial support from ANPCyT (grants PICT 2016-0791, PICT 2018-1509 and PICT 2019-0371), CONICET (grant PIP 11220150100506), from SeCyT-UNCuyo (grant 06/C603), from Italian Ministry for University and Research through PRIN Project No. 2017Z8TS5B and 20207ZXT4Z.
HPOC is supported by the Marie Sk{\l}odowska-Curie individual fellowship Grant agreement SUPERDYN No. 893743.
\end{acknowledgments}

\appendix
\section{Example of Dynamics for DOS-driving}\label{app:example-dynamics-dos}
%%%%%%%%%%%%%%%%%%%%%%%%%%%%%%%%%%%%%%%%%%%%%%%%%%%%%
\begin{figure}[h]
\includegraphics[width=0.5\textwidth]{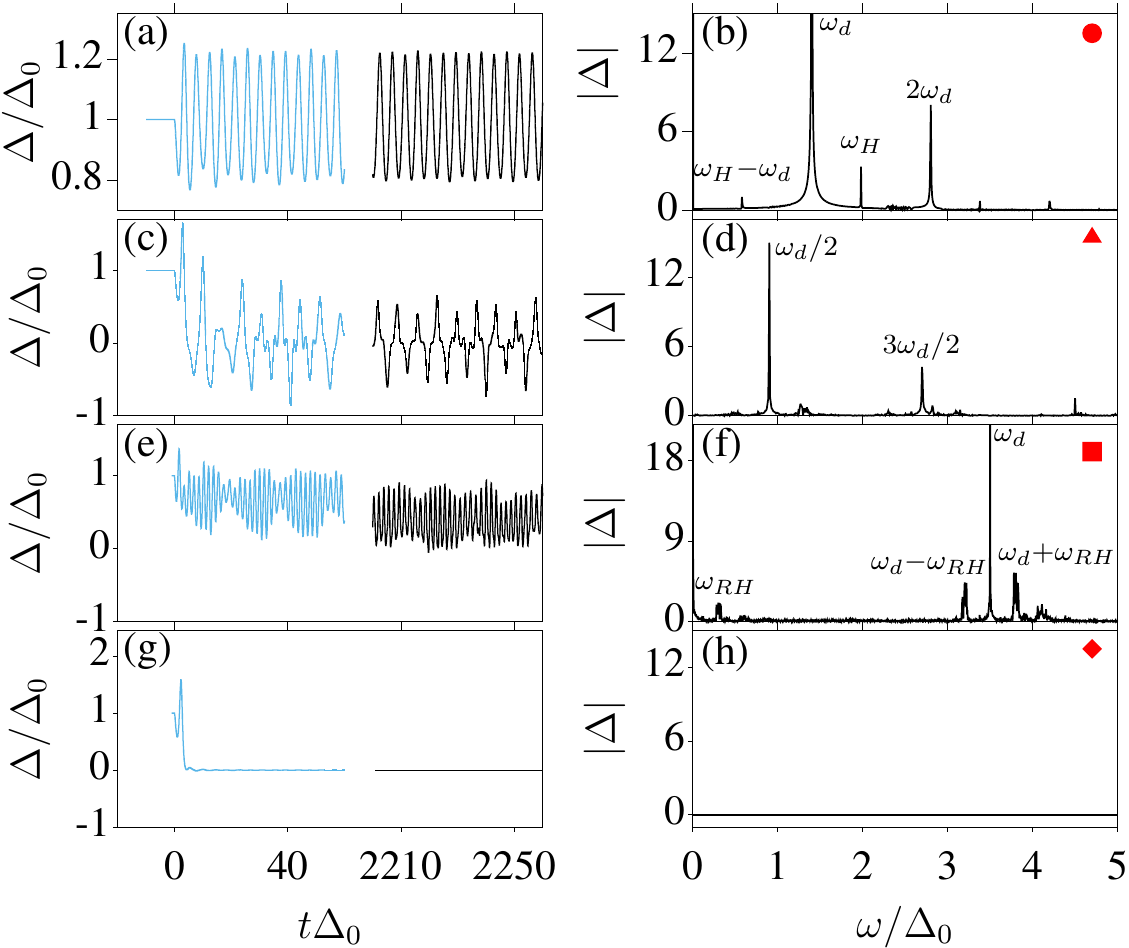}
\caption{(Color online) The same of Fig. 2 for the parameters indicated in the panel (b) of Fig. \ref{fig:phased}.} 
\label{fig:DOSdyn}
\end{figure}
%%%%%%%%%%%%%%%%%%%%%%%%%%%%%%%%%%%%%%%%%%%%%%%%%%%%%%%%



\end{document}